\documentclass[journal]{IEEEtran}
\usepackage{amsmath,amsfonts}
\usepackage{algorithmic}
\usepackage{algorithm}
\usepackage{array}
\usepackage[caption=false,font=normalsize,labelfont=sf,textfont=sf]{subfig}
\usepackage{textcomp}
\usepackage{multirow}
\usepackage{stfloats}
\usepackage{amssymb}
\usepackage{xcolor}
\usepackage{url}
\usepackage{verbatim}
\usepackage{graphicx}
\usepackage{cite}
\usepackage[table]{xcolor}
\usepackage{colortbl}
\usepackage{enumitem}
\usepackage{booktabs}
\usepackage[most]{tcolorbox}

\newtcolorbox{summarybox}{
    colback=blue!5,
    colframe=blue!60!black,
    boxrule=0.8pt,
    arc=3pt,
    left=6pt,
    right=6pt,
    top=6pt,
    bottom=6pt,
    before skip=10pt,
    after skip=10pt
}

\definecolor{improve}{RGB}{220,245,220}  % light green
\definecolor{degrade}{RGB}{245,220,220}  % light red

\newcommand{\relpos}[1]{\cellcolor{degrade}{#1}}
\newcommand{\relneg}[1]{\cellcolor{improve}{#1}}

\newcommand{\rowmax}[1]{\cellcolor{gray!15}{#1}}

\IEEEoverridecommandlockouts
\bstctlcite{IEEEexample:BSTcontrol}

\begin{document}

\bstctlcite{IEEEexample:BSTcontrol}

\title{Cyclostationarity Analysis as a Complement to Self-Supervised Representations for Speech Deepfake Detection}

\author{Cemal Hanil\c{c}i, Md. Sahidullah, Tomi H. Kinnunen
\thanks{C. Hanil\c{c}i is with Bursa Technical University, Bursa, Turkey. (e-mail: cemal.hanilci@btu.edu.tr). Md Sahidullah is with TCG CREST, Kolkata, India (e-mail: md.sahidullah@tcgcrest.org). T. Kinnunen is with the School of Computing, University of Eastern Finland, FI-80101 Joensuu, Finland (e-mail: tomi.kinnunen@uef.fi).}}

\markboth{Journal of \LaTeX\ Class Files,~Vol.~14, No.~8, August~2021}%
{Shell \MakeLowercase{\textit{et al.}}: A Sample Article Using IEEEtran.cls for IEEE Journals}


\maketitle

\begin{abstract}
Speech deepfake detection (SDD) is essential for maintaining trust in voice-driven technologies and digital media. Although recent SDD systems increasingly rely on self-supervised learning (SSL) representations that capture rich contextual information, complementary signal-driven acoustic features remain important for modeling fine-grained structural properties of speech. Most existing acoustic front ends are based on time-frequency representations, which do not fully exploit higher-order spectral dependencies inherent in speech signals. We introduce a cyclostationarity-inspired acoustic feature extraction framework for SDD based on spectral correlation density (SCD). The proposed features model periodic statistical structures in speech by capturing spectral correlations between frequency components. In particular, we investigate whether cyclostationary representations provide complementary information beyond conventional time-frequency features and modern SSL embeddings. To this end, we introduce temporally structured SCD representations that preserve the evolution of spectral and cyclic-frequency correlations over time. Their effectiveness is evaluated using multiple SDD architectures, including conventional neural networks, SSL-based embedding systems, and hybrid fusion models. Experiments on ASVspoof 2019 LA, ASVspoof 2021 DF, and ASVspoof 5 demonstrate that higher-order spectral correlations provide complementary information to SSL embeddings in several challenging conditions. In particular, fusion of SSL and SCD embeddings reduces the equal error rate on ASVspoof 2019 LA from 8.28\% to 0.98\%, and from 15.81\% to 14.80\% on the challenging ASVspoof 5 dataset. These findings establish higher-order cyclostationary spectral correlations as a complementary source of information for modern SDD, while providing new signal-processing insight into the statistical differences between natural and synthetic speech. 
\end{abstract}

\begin{IEEEkeywords}
ASVspoof, Spectral correlation density, Spoofing countermeasures
\end{IEEEkeywords}

\section{Introduction}

\IEEEPARstart{S}{peech} deepfake detection (SDD), also known as \emph{spoofing detection}~\cite{wu2015spoofing}, is the task of classifying a given utterance either as real (bonafide) or fake (spoof). The latter includes utterances generated using text-to-speech, voice conversion, adversarial attacks (or other similar means). SDD is increasingly important in audio forensics, secure voice communications, and maintaining trust in online audio content~\cite{spoofing_forensics, delgado2025deepfakevoicedetection}.

As with any classification task, the choice of feature representation is critical for SDD. Most SDD systems rely on neural network models, such as
\emph{convolutional neural networks} (CNNs)~\cite{sahidullah2019introduction,nautsch2021asvspoof, Res2Net50_ASVspoof} that excel in learning cues relevant to SDD from a fixed \emph{time-frequency} (TF) representation. Though the most common TF representation is the \emph{short-term Fourier transform} (STFT), numerous alternatives~\cite{todisco2017constant,sahidullah2015comparison}  have been considered~\cite{nautsch2021asvspoof}. An alternative to these hand-crafted feature representations are \emph{end-to-end} models (e.g. RawNet and AASIST) where the network implicitly learns a TF-like feature representation from the raw speech in a fully data-driven manner~\cite{9456037,tak2021end}. Despite typically outperforming fixed signal processing-based front-ends, fully learned front-end representations come with an increase in run-time and memory overhead. Moreover, since these representations are jointly optimized with the downstream task, they may capture dataset-specific characteristics. This motivates continued investigation of signal-processing features that provide information complementary, rather than redundant, to learned representations. 

Recently, \emph{self-supervised learning} (SSL) models have demonstrated strong performance for SDD by learning high-level speech representations directly from large-scale unlabeled speech corpora~\cite{wang22_odyssey,layerwise_ssl, layerwise_ssl2}. SSL-derived features provide rich contextual and linguistic information and have become a strong baseline for modern SDD. 
Nonetheless, being optimized for data-driven objectives rather than explicit signal-processing principles, they may not capture all higher-order statistical properties of speech. This motivates investigating whether theoretically motivated signal-processing representations can provide complementary information beyond that already captured by SSL features. 

In this study, we put forward a fixed TF analysis approach motivated by stochastic signal processing principles, particularly the concept of cyclostationarity. Although \emph{cyclostationarity} has a well-established theoretical foundation and clear relevance to the spectral analysis of speech signals, its use in SDD has received limited attention so far. While a familiar example of cyclostationarity is a signal produced by a known, periodic mechanism \cite{Gardner2006}, any signal that exhibits hidden periodicity or repetitive patterns is also considered cyclostationary \cite{Antoni2007}. These latter signals are typically aperiodic in time with random waveforms, but their statistical characteristics are periodic (or almost periodic) \cite{GardnerBook, Gardner1986}. 

Cyclostationarity is pertinent to communication signals \cite{Gardner2006,Antoni2007}, where periodicity arises from modulation, multiplexing, or coding. Speech can also be viewed as a communication signal and modeled as a cyclostationary random process driven by a nearly-periodic excitation signal during voiced speech production. Further, \emph{sinusoidal} modeling of speech~\cite{SpeechAMFM,SpeechAMFM2,Zeng2293} represents speech using a small number of modulated sinusoids, further supporting the relevance of cyclostationary analysis. Consequently, cyclostationary signal processing (CSP) techniques have previously been applied in speech modeling and analysis. For instance, \cite{paliwal97_eurospeech} proposed a linear prediction analysis based on a cyclic autocorrelation function, and \cite{patent_asr} demonstrated the use of cyclostationarity in speech enhancement. More recently, cyclostationary spectral analysis was successfully used for speech emotion recognition in \cite{SCD_Emotion}. Recent study also indicates that voiced speech is more accurately characterized as a cyclostationary process than as a wide-sense stationary process~\cite{bologni2024harmonics}.

Cyclostationarity manifests as spectral correlations between frequency components, which are captured by the spectral correlation density (SCD). SCD is a powerful analysis tool widely used in communication signal classification, particularly for distinguishing signals generated by different modulation schemes~\cite{SCDModulation}. It characterizes the cross-spectral density between frequency-shifted versions of a signal~\cite{gardner1986measurement}. Rather than replacing modern learned representations, in this work, we aim to investigate whether SCD-derived higher-order spectral correlations contribute additional discriminative information to conventional TF representations and SSL embeddings. Establishing such complementarity would not only improve SDD performance but also provide insight into the statistical properties of synthetic speech that remain underrepresented by current data-driven representations. Hence, the main research question we address is not whether cyclostationary analysis can replace existing learned representations, but rather to what extent higher-order spectral correlations provide an independent source of discriminative information that complements modern SSL representations.

Building on this motivation, the contributions of this work are threefold:
\begin{itemize}
    \item We systematically investigate higher-order spectral correlations in modern synthetic speech and demonstrate that cyclostationary analysis captures discriminative signal characteristics complementary to conventional time-frequency representations and SSL embeddings.
    \item We investigate practical temporally structured SCD representations that enable incorporation of cyclostationary information into modern deepfake detection systems.
    \item We systematically evaluate the proposed SCD representations in combination with conventional and SSL feature embeddings across multiple datasets, spoofing scenarios, and countermeasure architectures, demonstrating that higher-order spectral correlations provide complementary discriminative information beyond conventional time-frequency representations and learned speech representations. 
\end{itemize}

\section{Spectral Correlation Density}
\label{sec:scd}

\subsection{STFT Representation of Speech Signals}
The conventional spectrogram is based on the STFT, where a discrete-time speech signal $s[n]$ is divided into short, overlapping frames $x[n,t]$, each consisting of $N$ samples indexed by $n = 0, 1, \ldots, N-1$. Here, $t = 0, 1, \ldots, T-1$ denotes the frame index. The STFT is defined as,
\begin{equation}
    X(t,f) = \sum_{n=0}^{N-1} \underbrace{s[tR + n]w[n]}_{\equiv \; x[n,t]} e^{-j2\pi \frac{f}{f_s} n},
    \label{eq:stft}
\end{equation}
where $w[n]$ is a data-tapering window of length $N$, and $R$ is the hop size, resulting in an overlap of $N - R$ samples between adjacent frames. The variable $f \in [0, \tfrac{f_s}{2}]$ denotes the continuous frequency in Hz, where $f_s$ is the sampling rate. The frequency variable $f$ is typically discretized, leading to practical computation through the discrete Fourier transform (DFT). The squared magnitude of the STFT is popularly known as the \emph{spectrogram}.

\subsection{Cyclostationary Signal Processing}
Consider a discrete-time \emph{wide-sense stationary} (WSS) cyclostationary random process, denoted by $\{x[n, t] \in \mathbb{R}: n, t \in \mathbb{Z}\}$\footnote{Here, $\mathbb{R}$ and $\mathbb{Z}$ represent the fields of real and integer numbers, respectively. To align with the definition of a spectrogram, the random process is defined as $x[n, t]$, where $n$ represents the time variable and $t$ denotes the realizations of the random process for each element of the sample space.}, with a period $N_0 \in \mathbb{Z}$. Under the WSS assumption, $\mathbb{E}\{x[n, t]\} = \mathbb{E}\{x[n+N_0, t]\}$, where $\mathbb{E}\{\cdot\}$ denotes expectation. 
The \emph{autocorrelation function} of the process is defined as~\cite{Antoni2007},
\begin{equation} R_x[n, \tau] = \mathbb{E}\left\lbrace x\left[n+\frac{\tau}{2}, t \right] x\left[n-\frac{\tau}{2}, t\right] \right\rbrace, \label{eq:autocorrelation}
\end{equation}
and the process is said to exhibit \emph{second-order} cyclostationarity (assumed here) if $R_x[n, \tau] = R_x[n+N_0, \tau]$~\cite{Gardner2006, Antoni2007}. In \eqref{eq:autocorrelation}, $\tau \in \mathbb{Z}$ denotes the discrete-time lag variable (in samples), representing the separation between the two signal instances in the autocorrelation function. Speech signals can be viewed as realizations of such cyclostationary processes. In particular, the autocorrelation function of a voiced speech frame is (approximately) periodic ($N_0$ being analogous to the fundamental period of voiced excitation)~\cite[Ch.4]{RabinerBook}. 

Note that $R_x[n, \tau\!=\!0]=\mathbb{E}\left\lbrace x^2[n,t]\right\rbrace$ corresponds to the energy of the $n$th sample of the signal $x[n, t]$. Since $R_x[n, \tau]$ is periodic in $n$, it can be expanded as a Fourier series:

\begin{equation}
R_x[n,\tau] = \sum_{\alpha \in \mathcal{A}} r_x[\tau, \alpha] e^{j2\pi \alpha n},
\label{eq:autocorrelation_fs}
\end{equation}
where $\alpha$ is known as a \textbf{cyclic frequency} (CF) variable, with $\mathcal{A} = \{\alpha_0, \ldots, \alpha_{\text{max}}\}$ denoting its possible values. The Fourier expansion coefficients, denoted by 
$r_x[\tau, \alpha]$, are defined by
\begin{equation}
r_x[\tau, \alpha]  \triangleq \lim_{N\rightarrow \infty} \frac{1}{2N+1} \sum_{n=-N}^{N} R_x[n,\tau] e^{-j2\pi \alpha n}.
\label{eq:cyclic_autocorrelation}
\end{equation} 
For a given CF $\alpha$, $r_x[\tau,\alpha]$ is a function of a time-lag variable $\tau$. This \emph{cyclic autocorrelation function}~\cite{GardnerBook,Gardner1986,Gardner2006} can be computed directly from the $N$ samples in a frame (speech frame $x[n,t]$ is assumed to be zero for $n\geq N$) using:
\begin{equation}
    r_x[\tau, \alpha] = e^{j\pi \alpha \tau} \frac{1}{N} \sum_{n=\tau}^{N-1} x[n,t]x[n-\tau, t] e^{-j2\pi \alpha n T_s},
    \label{eq:caf_discrete_time_v1}
\end{equation}

where $T_s:=1/f_s$ is the sampling period of the signal. Note that for $\alpha=0$, the cyclic autocorrelation function becomes the conventional (non-cyclic) autocorrelation function of the frame $x[n,t]$. Note also that although $\tau$ must be an even integer in \eqref{eq:autocorrelation}, \eqref{eq:caf_discrete_time_v1} allows the use of odd values.

\subsection{Spectral Correlation Density and Its Properties}
\label{sec:SCD}
The \textbf{spectral correlation density} (SCD) is defined as the correlation of spectral components separated by $\alpha$ units. It is a complex-valued function of $f$ and $\alpha$ and computed as the Fourier transform of $r_x[\tau, \alpha]$~\cite{GardnerBook, Gardner2006, Antoni2007}:
\begin{equation}
\text{SCD}(f, \alpha) = \sum_{\tau=-\infty}^{\infty} r_x[\tau, \alpha] e^{-j2\pi f \tau}.
\label{eq:scd}
\end{equation}

For the zero CF ($\alpha = 0$), SCD becomes the power spectral density (PSD) function of the frame $x[n, t]$, by definition---PSD is the Fourier transform of the autocorrelation function~\cite[Ch.3]{HayesBook}. 

The direct definition of SCD in \eqref{eq:scd} requires calculating the cyclic autocorrelation function $r_x[\tau,\alpha]$ for each $\alpha \in \mathcal{A}$ using \eqref{eq:caf_discrete_time_v1}, followed by a Fourier transform, which is computationally expensive. To address this, various more efficient estimators have been developed. We employ the \emph{time-smoothed average periodogram} approach~\cite{Roberts1991,Gardner1986, TimeAveragedSCD}, an extension of Welch's PSD estimator~\cite{WelchPSD}. It computes SCD using the spectral components of the frame, shifted by $\pm \frac{\alpha}{2}$ in frequency~\cite{Gardner1986, Roberts1991, Gardner1991, isma2010}.  It utilizes the \emph{frequency-shifting property} of the Fourier transform, namely,
\begin{equation}
X\left(f \pm \frac{\alpha}{2}, t \right) = \sum_{n=0}^{N-1} x[n,t]e^{\pm j \pi \alpha n} e^{-j2\pi f n},
\label{eq:fshifted_Spectra}
\end{equation}
to compute spectral correlation (SC) of frame $t$ at CF $\alpha$ as,
\begin{equation}
\text{SC}(f,\alpha, t) = X\left(f - \frac{\alpha}{2}, t \right) X^{*}\left(f + \frac{\alpha}{2}, t \right),
\label{eq:framewise_SC}
\end{equation} 
where $^{*}$ stands for the complex conjugate. The SCD of the signal is obtained by time averaging the SCs of all frames:
\begin{equation}
\text{SCD}(f,\alpha) = \frac{1}{T}\sum_{t=1}^T \text{SC}(f,\alpha,t),
\label{eq:smoothed_periodogram}
\end{equation}
which serves as an empirical estimator of \eqref{eq:scd}. As the total number of frames ($T$) approaches infinity, the two converge and become equal~\cite{gardner1986measurement, Roberts1991}. The estimator in \eqref{eq:smoothed_periodogram} is generally not strictly unbiased due to finite-window effects. However, as shown in \cite{BOUSTANY2005166}, with appropriate window selection and sufficient overlap between adjacent frames, the estimator becomes asymptotically unbiased. Its variance decreases with the number of averaged frames and is proportional to the product of power spectral densities at frequencies $f \pm \frac{\alpha}{2}$, making it a consistent estimator. In this work, we use a Hamming window with overlap, a usual choice with well-documented bias-variance trade-off, as shown in \cite{BOUSTANY2005166}.

SCD in~\eqref{eq:smoothed_periodogram} measures the time-averaged correlation between the spectral components of a signal that are spaced apart by the CF $\alpha$. It represents the distribution of the signal's power with respect to the acoustic frequency $f$ and the CF $\alpha$~\cite{Antoni2007}. Unlike the traditional STFT spectrogram, the SCD is a function of two frequency variables, $f$ and $\alpha$, which allows it to capture the spectral features associated with the cyclostationary properties of the signal. Another important interpretation of the SCD is that a non-zero $\text{SCD}(f,\alpha)$ indicates spectral correlation associated with modulation at cyclic frequency $\alpha$ around carrier frequency $f$, thereby jointly characterizing modulation and carrier frequencies~\cite{Antoni2007}.

For real-valued signals, including speech, SCD satisfies specific conjugate symmetries~\cite{Roberts1991}: $\text{SCD}^{*}(f,\alpha)=\text{SCD}(-f,-\alpha)$ and $\text{SCD}(f,-\alpha) = \text{SCD}^{*}(f,\alpha)= \text{SCD}^{*}(-f,\alpha)$. It is therefore sufficient to estimate $\text{SCD}(f,\alpha)$ only in one quadrant of the \emph{bifrequency plane} defined by $(f,\alpha)$.

The range of the CF variable, $\alpha$, is inherently limited by the sampling rate $f_s$. In discrete-time analysis, the spectral components involved in \eqref{eq:framewise_SC} must remain within the Nyquist interval. Therefore, the maximum admissible CF satisfies $\alpha_{\max} \leq f_s$. This constraint ensures that the shifted frequency components $f \pm \frac{\alpha}{2}$ lie within the observable frequency range without incurring aliasing. Combined with the conjugate symmetry properties discussed above, it is sufficient to consider $\alpha \in [0, \alpha_{\max}]$, where $\alpha_{\max}=f_s$.

Synthetic speech signals generated by modern TTS systems may exhibit subtle periodicity-related artifacts, which can arise from modeling biases or overfitting effects in neural vocoders \cite{PeriodicityArtifacts1}, as well as from unnatural modulation structures introduced by frame-wise or block-wise conversion of acoustic features into audio waveforms \cite{modulation_artifacts}. Such artifacts are expected to differ from those typically observed in bonafide human speech. While these effects may remain difficult to discern in time-domain waveforms or conventional STFT-based spectrogram representations, they can be more readily characterized using the SCD function. Therefore, before presenting experimental results, the following two sections establish a foundation for the subsequent analysis by demonstrating how the SCD function reveals hidden periodicities and modulation structures critical for distinguishing bonafide (genuine) speech from spoofed utterances.

\section{SCD of Modulated Signals}
To establish an intuitive understanding of the properties of the SCD function, we first analyze synthetic modulated noise and sinusoidal signals. These controlled and application-independent 
examples serve to demonstrate how SCD uncovers latent modulation structures missed by STFT spectrograms. Thus, we demonstrate known SCD properties; readers familiar with SCD may skip to Section~\ref{sec:SCD-vocoded-speech}.

We consider the following four modulated signals: 
\begin{itemize}[noitemsep, topsep=0pt]
\item \emph{White noise amplitude-modulated by a sinusoidal carrier}:
\begin{equation}
x_1[n] = w[n] \cdot \cos(2\pi f_c n/T_s),
\label{eq:ex1_1}
\end{equation}
where $w[n]$ is zero-mean white noise with unit variance and $f_c=100$ Hz is the carrier frequency. 
\item \emph{Mixture of two modulated white noise signals}:
\begin{equation}
x_2[n] = w_1[n] \cdot \cos(2\pi f_{c_1} n/T_s) + w_2[n] \cdot \cos(2\pi f_{c_2} n/T_s),
\label{eq:ex1_2}
\end{equation}
where $w_1[n],w_2[n]\!\sim\!\mathcal{N}(0,1)$ are independent white noise processes, and $f_{c_1}\!=\!100$ Hz and $f_{c_2}\!=\!250$ Hz.
\item \emph{Sinusoidal amplitude-modulated signal}:
\begin{equation}
x_3[n] =  \cos(2\pi f_{0} n/T_s) \cdot \cos(2\pi f_{c} n/T_s), 
\label{eq:ex1_3}
\end{equation}
where $f_0 \ll f_c$. Here, the modulation frequency is $f_m\!=\!150$ Hz and the carrier frequency is $f_c\!=\!500$ Hz.
\item \emph{Amplitude-modulated sinusoid embedded in additive noise}:
\begin{equation}
x_4[n] = x_3[n] + w[n],
\label{eq:ex1_4}
\end{equation}
where $w[n]$ is zero-mean white noise with unit variance.
\end{itemize}

\begin{figure*}
\vspace{-3mm}
\centering
\includegraphics[scale=0.22]{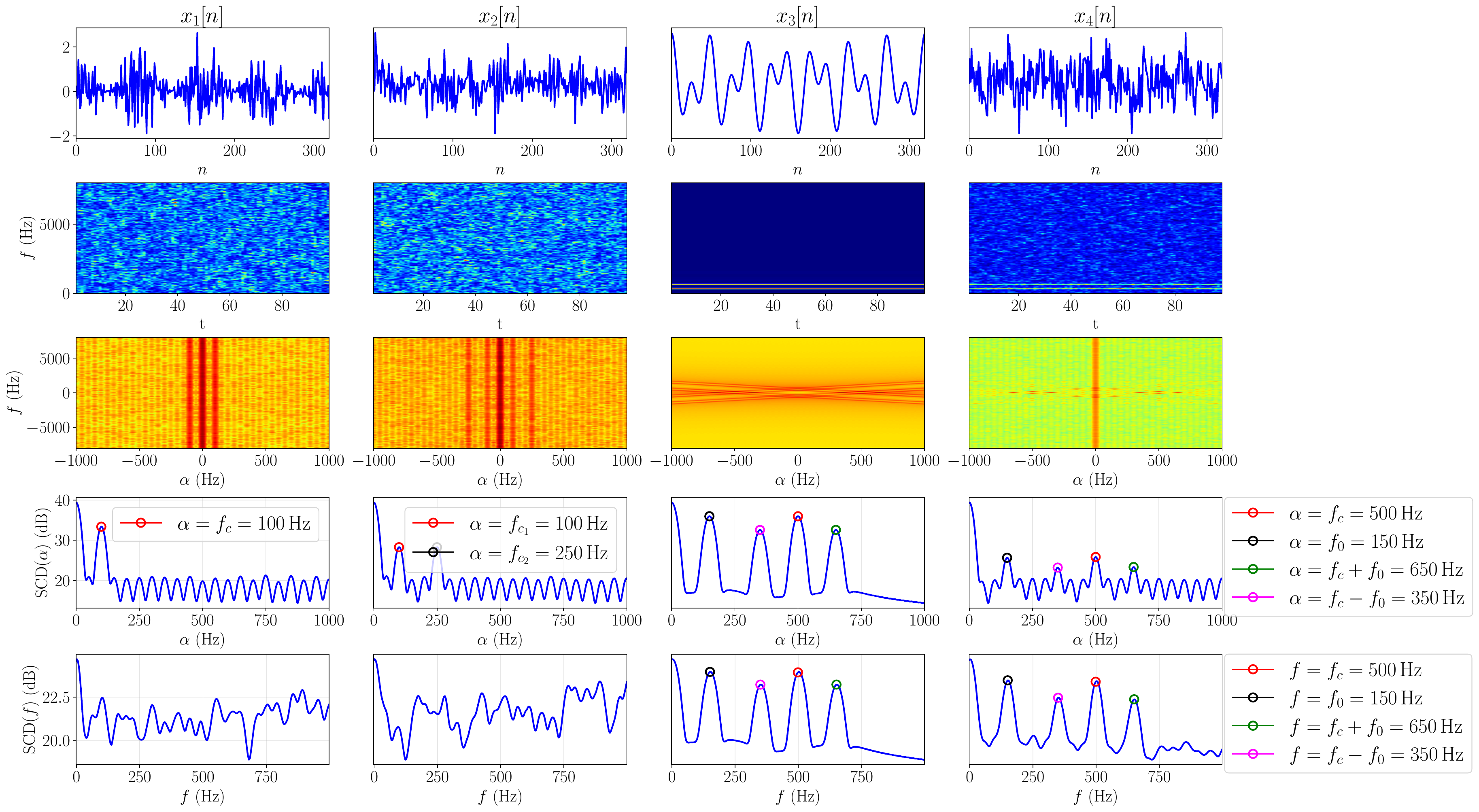}
\vspace{-3mm}
\caption{Four synthetic signals defined in \eqref{eq:ex1_1}–\eqref{eq:ex1_4}. From top to bottom: waveform, STFT spectrogram, SCD($f,\alpha$), averaged SCD($\alpha$), and averaged SCD($f$). One-dimensional plots are shown over 0–1000 Hz for clarity.}
\label{fig:modulated_signals_visualizations}
\end{figure*}

All signals are one second long and are sampled at $f_s = 16$ kHz. Fig.~\ref{fig:modulated_signals_visualizations} visualizes the waveforms, the STFT spectrograms, the conventional SCD representations $\text{SCD}(f,\alpha)$, the one-dimensional $\text{SCD}(\alpha)$ obtained by averaging $\text{SCD}(f,\alpha)$ over spectral frequency $f$, and the one-dimensional $\text{SCD}(f)$ obtained by marginalizing over the CF $\alpha$. 

For $x_1[n]$, $x_2[n]$, and $x_4[n]$, the waveforms appear noise-like due to the dominant stochastic white noise, which masks the underlying periodic modulation. In contrast, $x_3[n]$ displays a high-frequency carrier modulated by a low-frequency envelope, resulting in two spectral components at $f_c \pm f_m = 350$ Hz and $650$ Hz\footnote{This is derived using the trigonometric identity $\cos(A) \cdot \cos(B) = \frac{1}{2} \cos\left(A+B\right) + \frac{1}{2} \cos\left(A-B\right)$}. Identifying these sinusoids directly from the waveforms is not easy, however. Similarly, the STFT spectrograms reveal little modulation information for $x_1[n]$, $x_2[n]$, and $x_4[n]$, while the carrier components of $x_3[n]$ remain visible without explicitly separating carrier and modulation frequencies.

In contrast, the conventional SCD representation provides clear insight into the cyclostationary structure of the signals. For $x_1[n]$,  we observe prominent vertical ridges at cyclic frequencies $\alpha = \pm 100$ Hz, extending across all spectral frequencies. This indicates strong spectral correlation induced by the carrier of the modulating sinusoid.  Similarly, for $x_2[n]$, the apparent vertical ridges at $\alpha\!=\!f_{c_1}\!=\! \pm 100$ Hz and $\alpha\!=\!f_{c_2}\!=\!\pm 250$ Hz indicate persistent correlations between spectral components separated by the respective carrier frequencies. The SCD of $x_3[n]$, in turn, reveals diagonal structures corresponding to the interaction between the carrier and the modulation components. Finally, despite the presence of additive noise, these diagonal patterns remain clearly visible in the SCD of $x_4[n]$, highlighting the resilience of cyclostationary analysis to noise. For the noisy signal $x_4[n]$, the additive white Gaussian noise results in an SNR of approximately $-6$~dB\footnote{Using the definition of $x_3[n]$ in~\eqref{eq:ex1_3} and assuming $w[n]\sim\mathcal{N}(0,1)$, the signal power becomes $P_{x_3}=1/4$ while the noise power is $P_w=1$. This gives $\mathrm{SNR}=10\log_{10}(1/4)\approx -6.02$~dB.}. Additional experiments under varying SNR conditions showed that additive noise primarily increases the background floor of the SCD representation while preserving the locations of the modulation-, carrier-, and sideband-related peaks. The ridge at $\alpha=0$ for $x_1[n],\: x_2[n]$ and $x_4[n]$ consistently represents the conventional PSD.

The one-dimensional averaged SCD representations further reveal the latent spectral correlations. For $x_1[n]$ and $x_2[n]$, we observe strong spectral correlations at their respective carrier frequencies. For $x_3[n]$, $\mathrm{SCD}(\alpha)$ displays distinct peaks at the modulation, carrier, and sideband frequencies. Even for $x_4[n]$, these characteristic peaks remain visible, underlining the noise robustness of SCD \cite{SCD_robustness_wgn1, SCD_robustness_wgn2}. Whereas $\mathrm{SCD}(f)$ for $x_1[n]$ and $x_2[n]$ does not directly reflect the carrier frequencies, for $x_3[n]$ and $x_4[n]$ we observe distinct peaks corresponding to the carrier, modulation and sideband frequencies. This behavior can be explained by~\eqref{eq:framewise_SC}, where $\mathrm{SC}(f,\alpha,t)$ measures the correlation between spectral components separated by $\alpha$. For noise-modulated signals ($x_1[n]$ and $x_2[n]$), these spectral components are broadband and largely uncorrelated, so the resulting $\mathrm{SCD}(f,\alpha)$ is spread over $f$ and does not yield dominant peaks after averaging over $\alpha$. In contrast, for $x_3[n]$ and $x_4[n]$, sinusoidal modulation produces discrete and deterministic spectral lines (e.g., at $f_c \pm f_0$) that are strongly correlated, resulting in localized peaks in $\mathrm{SCD}(f)$. This reveals that the conventional $\mathrm{SCD}(f,\alpha)$ representation encapsulates details about both the carriers and modulators. 
\begin{summarybox}
\textbf{Summary.}  
The analysis of modulated signals presented above highlights the following key points:
\begin{itemize}
    \item STFT fails to capture modulation structure in noise-dominated signals. 
    \item SCD recovers latent periodicities by revealing distinct peaks at modulation, carrier and sideband frequencies, even when this modulation structure is embedded in white noise.
    \item SCD remains robust under additive white noise corruption. 
    \item Vertical ridges in $\mathrm{SCD}(f,\alpha)$ indicate strong cyclostationary behaviour while diagonal structures are indicative of modulation-induced spectral correlations.
\end{itemize}
\end{summarybox}

These properties make SCD a useful tool in signal detection and classification, especially in noisy environments, where traditional time–frequency techniques face their limits. 

\section{SCD of Vocoded and Synthetic Speech}\label{sec:SCD-vocoded-speech}
With the intuition established from synthetic examples, we now turn to speech signals, whose modulation structure is considerably more complex. We focus on comparing the SCD representations of bonafide speech and speech synthesized by waveform generation models (vocoders) to investigate how vocoder-induced modifications of the natural excitation-filter structure are reflected in higher-order spectral correlations.

\subsection{Vocoder Models}
We analyze speech synthesized using both DSP-based and data-driven vocoders, beginning with classic \emph{sinusoidal}~\cite{McAulay_Sinusoidal_Modeling} and \emph{linear predictive coding} (LPC) vocoders, which model speech using periodic or quasi-periodic components. We estimate their parameters directly from bonafide speech and subsequently resynthesize an approximated waveform, enabling a controlled comparison of their SCDs.

The sinusoidal model \cite{McAulay_Sinusoidal_Modeling} approximates the samples of a speech frame $s[n]$ as a sum of $K$ sinusoids,
\begin{equation}
s[n] \approx \sum_{k=1}^K A_k \sin(\omega_k n + \phi_k),
\label{eq:sinusoidal_model}
\end{equation}

where $A_k$, $\omega_k$, and $\phi_k$ are the amplitude, angular frequency, and phase of the $k$th sinusoid, respectively. A speech frame can accordingly be interpreted as a cyclostationary signal consisting of multiple periodic components. We employ the classical spectral analysis-based sinusoidal parameter estimation method~\cite{McAulay_Sinusoidal_Modeling} where prominent peaks in the magnitude spectrum of a windowed speech frame are first detected and their corresponding frequencies, amplitudes and phases are extracted. These spectral peaks are then used to fit the sinusoidal model in~\eqref{eq:sinusoidal_model} to the bonafide speech signal.  

In contrast, LPC~\cite{Makhoul_LP} models speech as the output of an all-pole filter excited by either a periodic impulse train for voiced speech or white noise for unvoiced speech, and is widely used in standardized speech codecs such as CELP, MELP, and AMR. More recently, it has also been integrated with neural vocoders (e.g. LPCNet~\cite{LPCNet}) to improve synthesis quality. Understanding the cyclostationary characteristics of LPC-synthesized speech is therefore relevant for both legacy and modern TTS systems. In LPC, a speech sample $s[n]$ is predicted from its previous $p$ samples as $\hat{s}[n] = - \sum_{k=1}^p a_k s[n-k]$ where $\hat{s}[n]$ denotes the predicted sample, $a_k$ are the predictor coefficients and $p$ is the predictor order. The coefficients $a_k,\: k=1,\ldots p$ are obtained by minimizing the mean-squared prediction error of the residual signal $e[n] = s[n]-\hat{s}[n]$. The frequency response of the resulting all-pole synthesis filter is  $H(z) = 1/\big(1 + \sum_{k=1}^p a_k z^{-k}\big)$. The synthesis filter can then be excited with a suitable source signal to synthesize speech. 
 
While sinusoidal and LPC vocoders provide useful insight, thanks to their simplicity and transparency, vocoder implementations used in practical TTS and VC systems are more nuanced. Hence, we also include another widely used DSP-based vocoder called WORLD~\cite{World_Vocoder}. This lightweight vocoder decomposes speech into three sets of parameters: spectral envelope, fundamental frequency ($F_0$), and aperiodicity. These parameters are then used to reconstruct speech with improved temporal resolution and perceptual quality relative to sinusoidal and LPC vocoders.

In addition to the above DSP-based approaches, we further include a representative collection of modern neural vocoders: CARGAN \cite{cargan}, FarGAN \cite{Fargan}, HiFi-GAN \cite{hifigan}, LPCNet \cite{lpcnet2}, and WaveGAN \cite{wavegan}. They employ deep generative architectures to learn complex, data-driven mappings between acoustic features and speech waveforms. These neural vocoders perform end-to-end optimization, often achieving superior perceptual quality and increased temporal detail. By applying SCD analysis to their outputs, we aim to assess whether neural vocoders mitigate or introduce distinctive modulation artifacts and spectral redundancies, thereby providing insights into the detectability of contemporary speech deepfakes.  

\subsection{SCD Analysis of Vocoded Speech}

To ensure consistency, all synthetic speech signals were generated from a single utterance spoken by a male speaker (\texttt{mic\_M02\_si760.wav})\footnote{This bonafide speech signal and its synthesized counterparts generated using neural vocoders are publicly available at \url{https://ahmed-fau.github.io/fargan_demo/}, provided by the authors of \cite{Fargan}.} with a duration of approximately seven seconds. For sinusoidal modeling and LPC vocoders, the original (bonafide) speech was first segmented into 20~ms frames with 10~ms overlap and a Hamming window was applied to each frame. Model parameters---namely, $(A_k, \omega_k, \phi_k)$ for sinusoidal modeling and $a_k$ for LPC synthesis---were estimated on a frame-by-frame basis. For the LPC vocoder, a $20$th order LPC model ($p=20$) was estimated using the autocorrelation method \cite{Makhoul_LP}. Voiced frames were re-synthesized using a perfectly periodic impulse train, with the fundamental period estimated from the LP residual, whereas unvoiced frames were re-synthesized using white noise excitation. Voicing was determined on a frame-by-frame basis using an autocorrelation-based pitch analysis of the LPC residual as described in \cite{pitch_estimation_lpc_residual}. Finally, for both sinusoidal and LPC-based synthesis, the reconstructed frames were combined using overlap-and-add to produce the complete waveform.

\begin{figure*}
\vspace{-3mm}
\centering
\includegraphics[scale=0.25]{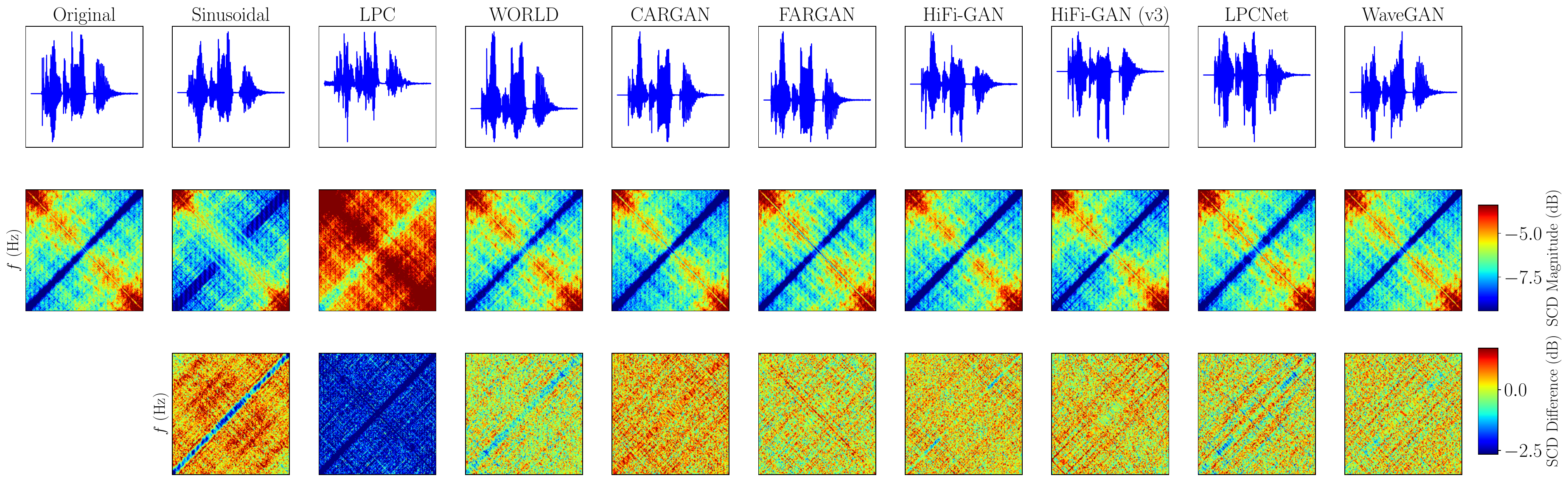}
\vspace{-3mm}
\caption{Comparison of bonafide and synthesized speech in the time and cyclostationary domains. \textbf{Top}: Waveforms of the original utterance and signals synthesized using sinusoidal modeling, LPC, WORLD, and neural vocoders. \textbf{Middle}: Corresponding $\mathrm{SCD}(f,\alpha)$ representations. \textbf{Bottom}: Difference maps, $\Delta \mathrm{SCD}(f,\alpha)=\mathrm{SCD}_{\mathrm{original}}(f,\alpha)-\mathrm{SCD}_{\mathrm{synthesized}}(f,\alpha)$, highlighting vocoder-induced artifacts. Due to symmetry of the SCD representation, only the first quadrant is shown. All synthesized signals were generated from the same bonafide utterance. For clarity, waveforms show only the first second, whereas SCDs are estimated from the full utterance.}
\vspace{-3mm}
\label{fig:vocoded_speech_examples}
\end{figure*}

Fig.~\ref{fig:vocoded_speech_examples} displays the waveforms and SCDs of bonafide and synthesized speech, along with SCD difference maps to highlight vocoder-induced artifacts in the cyclostationary domain. Whereas the waveform comparison indicates that all vocoders preserve the overall temporal structure, each vocoder introduces subtle deviations. 

The bonafide speech signal exhibits a highly structured pattern characterized by pronounced diagonal components, indicative of strong second-order cyclostationarity arising from the periodic excitation and harmonic structure. Whereas we observe similar diagonal structure for all vocoded signals, we also observe structural differences from the bonafide signal, providing a promising basis for distinguishing bonafide speech from vocoded speech.

Among the DSP vocoders, sinusoidal and LPC-based methods show substantial deviations from the SCD of bonafide speech. The LPC vocoder in particular exhibits a marked attenuation of structured cyclostationary components, as evidenced by the predominantly negative SCD difference (blue regions), suggesting a loss of periodic and harmonic dependencies due to the oversimplified excitation modeling. Sinusoidal modeling, while preserving certain diagonal structures, introduces excessive energy at specific cyclic frequencies, leading to pronounced artifacts visible as strong positive SCD differences. The WORLD vocoder demonstrates improved preservation of the global SCD structure over the LPC and sinusoidal vocoders. Nonetheless, its SCD difference map reveals widespread low-magnitude discrepancies across the full bifrequency plane. Hence, although WORLD captures harmonic organization more effectively, it still introduces subtle but systematic distortions in cyclostationary statistics.

The neural vocoders have SCDs that closely resemble those of the bonafide speech signal. Their SCDs retain the characteristic diagonal structure and symmetry observed in the bonafide signal, suggesting improved modeling of periodic excitation and temporal dependencies. Among these models, HiFi-GAN and HiFi-GAN (v3) display the smallest, most uniformly distributed SCD differences. LPCNet also performs competitively, although its SCD difference reveals localized artifacts at higher cyclic frequencies, potentially attributable to limitations of its hybrid parametric–neural excitation model.

WaveGAN and GAN-based autoregressive models (CARGAN and FARGAN) exhibit moderate SCD discrepancies, often manifested as fine-grained speckle-like patterns in the difference maps. These artifacts suggest that, while adversarial training enables realistic waveform generation in the time domain, subtle inconsistencies remain in higher-order temporal correlations captured by the SCD representation.

Overall, the SCD analysis reveals clear distinctions between bonafide and synthesized speech that are not always apparent from the time-domain waveforms alone. Traditional vocoders introduce pronounced and structured deviations in the cyclostationary domain, whereas modern neural vocoders substantially reduce—but do not entirely eliminate—these discrepancies. This indicates that the improved realism of neural vocoders further underscores the value of SCD-based analysis as a fine-grained analysis tool for characterizing synthesis artifacts and differentiating bonafide and synthetic speech in increasingly challenging scenarios.

\section{Temporal SCD Features}
The way SCD is used for feature extraction plays a critical role in its effectiveness. The conventional SCD representation defined in~\eqref{eq:smoothed_periodogram} averages the per-frame spectral correlation (SC) over the time (frame) axis. Beyond averaging over time, the SCD representation can also be formed by marginalizing over the frequency ($f$) or CF ($\alpha$) dimensions in~\eqref{eq:framewise_SC}. Incorporating temporal variations could enhance the utility of SCD in deepfake detection, as temporal dynamics are important in many speech classification tasks. To analyze this, we consider two alternative SCD-based feature representations obtained via marginalization:
\begin{equation}
\begin{aligned}
\text{SCD}_a(\alpha,t) & = \frac{1}{N_\text{fft}}\sum_{f} \text{SC}(f,\alpha,t)\\
\text{SCD}_b(f,t) & = \frac{1}{|\mathcal{A}|}\sum_{\alpha} \text{SC}(f,\alpha,t).
\label{eq:scda_scdb}
\end{aligned}
\end{equation} 
where $N_\text{fft}$ denotes the number of DFT frequency bins, $\mathcal{A}$ denotes the set of CFs used in the SC computation, and $|\mathcal{A}|$ denotes its cardinality.

\begin{figure*}[!t]
\vspace{-3mm}
\centerline{\includegraphics[scale=0.17]{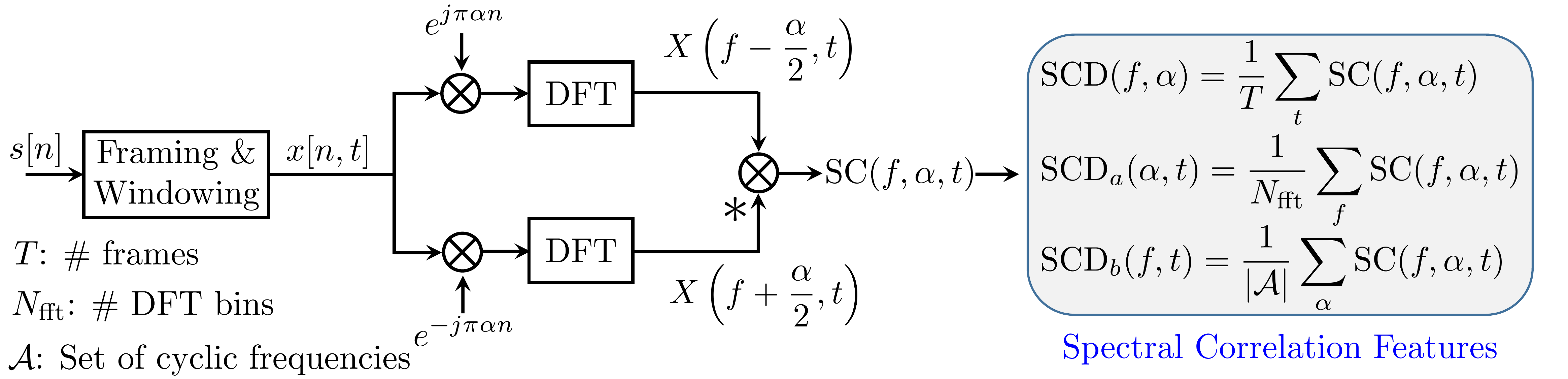}}
\vspace{-3mm}
\caption{Estimation of conventional and temporal SCD features ($\text{SCD}_a$, $\text{SCD}_b$) from the speech signal $s[n]$. $^{*}$ denotes complex conjugation.}
\vspace{-3mm}
\label{fig:SCD_steps}
\end{figure*}

The first representation, denoted by $\text{SCD}_a$, captures the SC variation over CF ($\alpha$) and time ($t$) by aggregating information across acoustic frequency bins ($f$), thus summarizing global cyclostationary behavior. Conversely, the $\text{SCD}_b$ representation captures the SC variation as a function of acoustic frequency ($f$) and time ($t$), emphasizing the cyclostationary properties while preserving the temporal structure. Notably, while STFT spectrograms and $\text{SCD}_b$ share dependencies on $f$ and $t$, $\text{SCD}_b$ uniquely encodes cyclostationary characteristics. The block diagram in Fig.~\ref{fig:SCD_steps} summarizes the three alternative feature extraction schemes.

\begin{figure}[!h]
\vspace{-3mm}
\centerline{\includegraphics[scale=0.27]{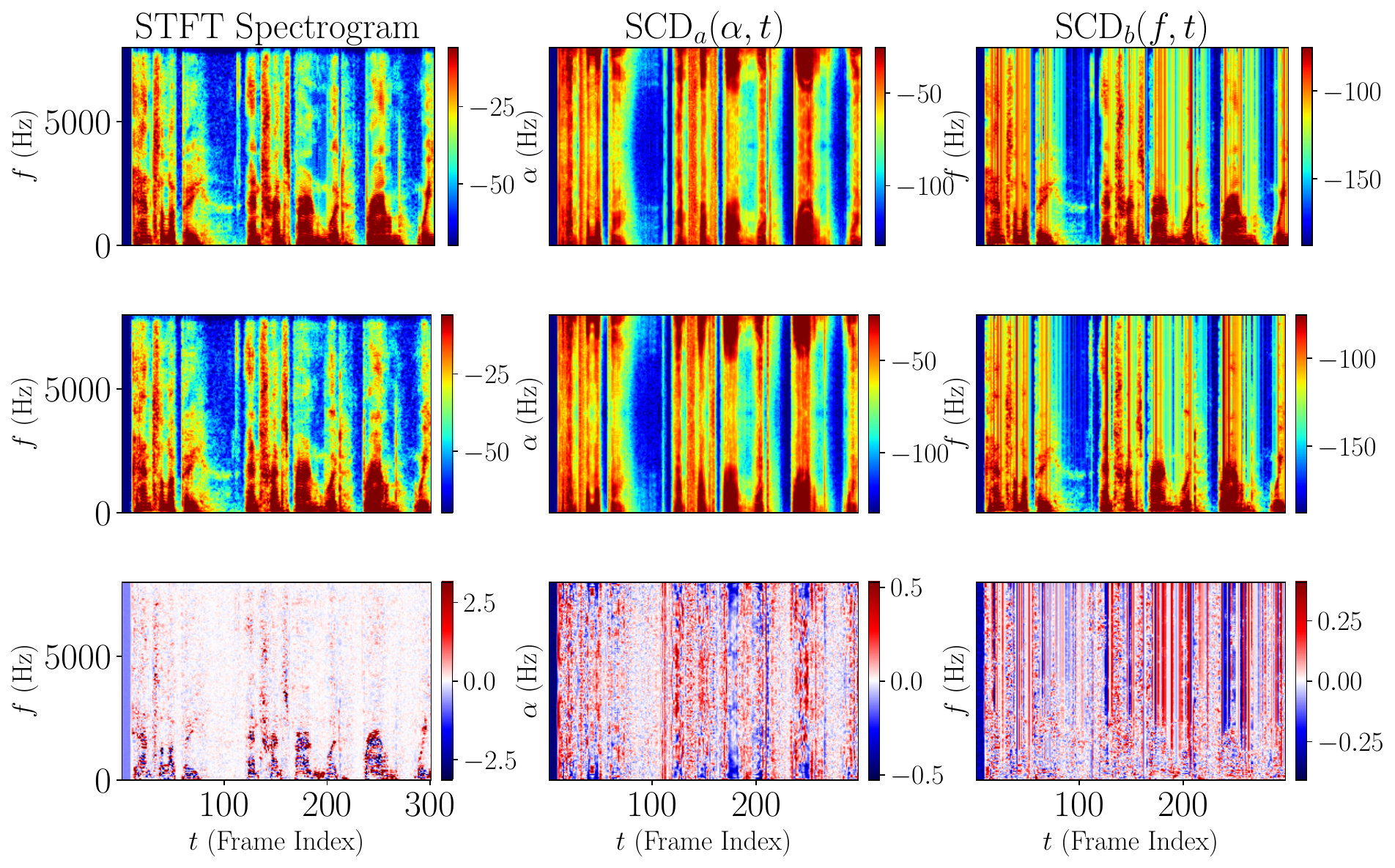}}
\vspace{-3mm}
\caption{Time-frequency representations of a bonafide speech signal and its HiFi-GAN (v3) synthetic counterpart. Rows correspond to the bonafide signal, the synthetic signal, and their relative difference maps, while columns show the STFT spectrogram, $\mathrm{SCD}_a(\alpha,t)$, and $\mathrm{SCD}_b(f,t)$ representations.}
\vspace{-3mm}
\label{fig:STFT_SCDa_SCDb}
\end{figure}

To provide further insight into the characteristics of the proposed representations and to highlight their differences from conventional STFT spectrogram features, Fig.~\ref{fig:STFT_SCDa_SCDb} visualizes the STFT spectrogram, $\mathrm{SCD}_a(\alpha,t)$, and $\mathrm{SCD}_b(f,t)$ computed from the same bonafide utterance and its synthetic HiFi-GAN (v3) counterpart used in Sect.~\ref{sec:SCD-vocoded-speech}. The first two rows show the feature representations of the bonafide and synthetic signals, respectively, while the third row presents relative difference maps that emphasize vocoder-induced deviations.

While the three representations (STFT, $\mathrm{SCD}_a$ and $\mathrm{SCD}_b$) of the bonafide and synthetic signals appear visually similar in the first two rows, the relative difference map of the STFT spectrogram reveals only weak and spatially diffuse deviations, indicating that these representations are largely preserved by the neural vocoder. In contrast, the relative difference maps of $\mathrm{SCD}_a(\alpha, t)$ and $\mathrm{SCD}_b(f,t)$ exhibit pronounced, temporally persistent vertical structures spanning a wide frequency range, suggesting systematic alterations in the SC patterns introduced by the synthesis process. Although finite-length cyclic spectral estimation is known to introduce spectral/cyclic leakage effects depending on the selected analysis
window~\cite{harris_window_functions,BOUSTANY2005166}, the observed
vertical structures are unlikely to originate solely from windowing
artifacts. In particular, these structures remain visible in the
relative difference maps computed between the bonafide and synthetic
signals, even though both signals were processed using the same window (Hamming), segment overlap, and estimation procedure. Moreover, similar patterns were consistently observed when Hanning and Blackman windows were used, although their prominence was slightly reduced due to the stronger sidelobe suppression of these windows. These observations suggest that the vertical structures are also associated with intrinsic differences in the temporal cyclic spectral organization of natural and synthetic speech signals introduced by the neural vocoder.
This observation suggests that SCD features are more sensitive to vocoder-induced artifacts than conventional time--frequency magnitude representations, providing intuition for the superior discriminative performance of SC-based features observed in the quantitative spoofing-detection experiments presented next.

\section{Experimental Setup}
\subsection{Datasets}
We evaluate SDD performance on three datasets of increasing difficulty: the logical access (LA) task of ASVspoof 2019 \cite{ASVspoof2019}, ASVspoof 2021 Deepfake (DF) \cite{ASVspoof2021}, and ASVspoof 5 \cite{ASVspoof5}. All speech signals in these datasets are sampled at 16~kHz, ensuring a consistent frequency resolution across the experiments.

\textbf{ASVspoof 2019 LA}~\cite{ASVspoof2019} is a widely used dataset consisting of high-quality spoofing attacks. Its training and development subsets contain spoofed utterances generated using the same six TTS and VC algorithms (A01–A06). These \emph{known attacks} are used for model training and hyperparameter tuning, respectively. The evaluation set includes spoofed utterances generated with thirteen unseen algorithms (A07–A18), referred to as \emph{unknown attacks}.

\textbf{ASVspoof 2021 DF} is more challenging and contains bonafide and spoofed utterances processed using lossy audio codecs such as MP3 and OGG \cite{ASVspoof2021}. Following the recommended protocol~\cite{ASVspoof2021}, models are trained and tuned using the ASVspoof 2019 LA training and development sets, while ASVspoof 2021 DF is used only for evaluation.

\textbf{ASVspoof 5}~\cite{ASVspoof5} is the latest edition in the ASVspoof series. Representing a much larger and even more challenging testbench for SDD, it is built from the MLS English corpus, which contains over 4000 speakers. The spoofed utterances were generated using the latest TTS and VC systems, validated to effectively deceive an ECAPA-TDNN-based~\cite{ecapa_tdnn} ASV system. Similar to ASVspoof 2021 DF, the evaluation data have been processed using lossy audio codecs. Consistent with prior protocols, the training and development subsets are used for model training and tuning, while the evaluation set is reserved for performance assessment. 

Previous studies have shown that the ASVspoof 2019 and 2021 datasets are highly biased toward silence regions, which can lead to overly optimistic and unrealistic performance estimates~\cite{muller21_asvspoof, asvspoof_silence}. To mitigate this issue, silence segments are removed from the training utterances in experiments involving these datasets, enabling a less biased evaluation of the proposed spectral correlation features and spoofing countermesure (CM) systems. 

All CM systems are trained using the corresponding training sets, without any additional data augmentation. This allows us to assess the inherent robustness of the individual features and ensures a fair comparison among the evaluated approaches. In addition, avoiding augmentation keeps the training pipeline lightweight and computationally efficient, facilitating reproducibility and scalability across datasets. 
\subsection{Feature Extraction} 
\label{sec:features}
In addition to the proposed SC-based features, we include a number of established front-ends as our baselines. This set includes LFCCs \cite{sahidullah2015comparison}, Mel-spectrograms \cite{subnetwork_asvspoof}, STFT spectrograms \cite{sahidullah2015comparison}, and CQT spectrograms \cite{todisco2017constant}. Except for CQT, all features are extracted using 25~ms Hamming windows, a 10~ms frame shift, and a 512-point DFT. CQT uses a 10~ms frame shift without fixed frame length.

\begin{itemize}[noitemsep, topsep=0pt]
\item \textbf{LFCCs} were computed by following the official baseline parameterization provided by the ASVspoof 2019 challenge \cite{ASVspoof2019}. They were extracted from a 20-channel filterbank spaced on a linear scale. The 20-dimensional static features were augmented with delta and double-delta coefficients, resulting in 60-dimensional features. 
\item \textbf{Mel-spectrograms} were extracted using an 80-channel Mel-scale filterbank, yielding 80-dimensional features.
\item \textbf{CQT spectrograms} were obtained with 84 bins per octave, resulting in 84-dimensional features.
\item \textbf{STFT spectrograms and SC-based features} used a 512-point DFT, yielding 257 frequency bins.
\end{itemize}

The selection of the above feature representations enables a comparison between compact representations that apply filterbank-based smoothing (LFCCs and Mel-spectrograms) or logarithmic binning (CQT spectrograms), and full representations (STFT spectrograms, $\text{SCD}$, $\text{SCD}_a$, and $\text{SCD}_b$).

For training, the number of frames per utterance was fixed to 400 by cropping or replicating frames as needed, while evaluation was performed on full-length utterances. The maximum CF parameter ($\alpha_\text{max}$) for the SC-based features was determined through preliminary experiments on the development set of the ASVspoof 2019 LA dataset with respect to the equal error rate (EER) criterion. Although $\alpha$ is theoretically bounded by the sampling frequency (i.e., $\alpha \leq f_s$), in practice smaller values were found to be more effective. Accordingly, $\alpha_\text{max}$ was set to 2000, 2500, and 500~Hz for the conventional SCD, $\text{SCD}_a$, and $\text{SCD}_b$ features, respectively.

The resulting feature dimensions are $F \times T$, where $F \in \{60, 80, 84, 257, 257, 257\}$ for LFCCs, Mel-spectrograms, CQT spectrograms, STFT spectrograms, $\text{SCD}_a$ and $\text{SCD}_b$, respectively. Here, $T$ denotes the number of frames (fixed to 400 during training). The conventional $\text{SCD}$ features, in turn, are of dimension $257\times 257$ where the first dimension spans the spectral frequency range $[0,\:\frac{f_s}{2}]$, while the second dimension spans the CF range $[0,\alpha_\text{max}]$.

% CM Details
\subsection{Speech Deepfake Detection Models} 
\label{sec:classifiers}

\begin{figure*}[!t]
\vspace{-3mm}
\centerline{\includegraphics[scale=0.11]{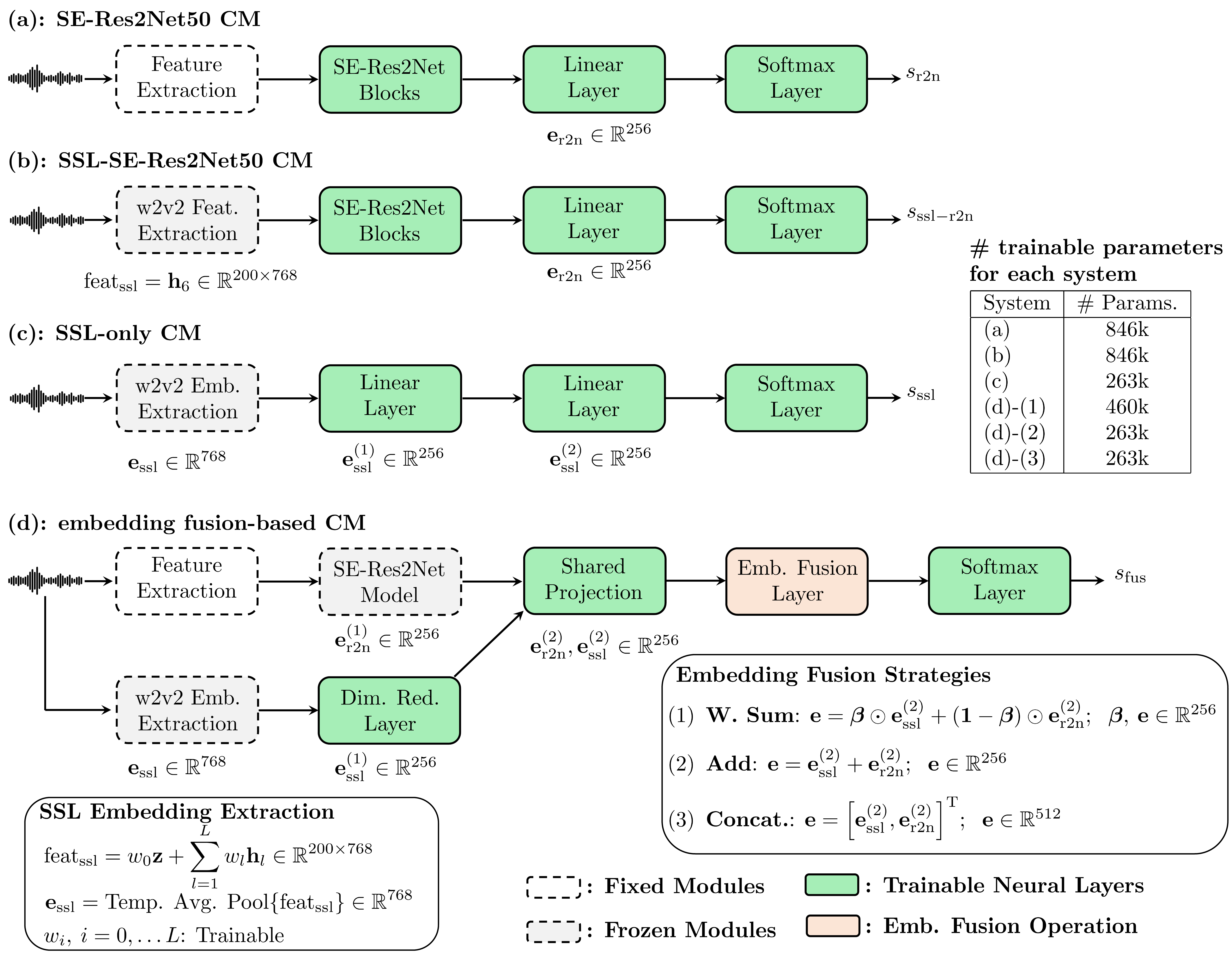}}
\vspace{-3mm}
\caption{Block diagrams of the four evaluated SDD CMs: (a) SE-Res2Net50 with handcrafted acoustic features; (b) SSL--SE-Res2Net50 with frozen Wav2Vec~2.0 frame-level features; (c) SSL-only with utterance-level Wav2Vec~2.0 embeddings; and (d) embedding fusion combining SE-Res2Net50 and SSL embeddings via weighted sum (W. Sum), addition (Add), or concatenation (Concat.). Dashed and solid blocks denote frozen and trainable modules, respectively. Intermediate dimensionalities are shown. The accompanying table reports trainable parameters (rounded to the nearest thousand), excluding the frozen SSL backbone.}
\vspace{-3mm}
\label{fig:cm_systems}
\end{figure*}

We consider four SDD models (spoofing countermeasures, CMs): 
\begin{itemize}[noitemsep, topsep=0pt]
\item a conventional \textbf{SE-Res2Net50-based CM} \cite{Res2Net50_ASVspoof}, 
\item an \textbf{SSL-SE-Res2Net50 CM},
\item an \textbf{SSL-only CM} based on Wav2Vec~2.0 embeddings, similar to the weighted layer-averaging approach in \cite{stourbe24_asvspoof} but employing Wav2Vec~2.0 \cite{wav2vec2} instead of WavLM, and
\item an \textbf{embedding fusion-based CM} that combines complementary SSL and convolutional representations.
\end{itemize}
The block diagrams of all systems and their parameter counts are shown in Fig.~\ref{fig:cm_systems}. The SE-Res2Net50-based CM serves as a strong and well-established convolutional baseline, with competitive performance particularly on the ASVspoof 2019 dataset \cite{Res2Net50_ASVspoof}. Moreover, its convolutional architecture is well-suited for processing two-dimensional representations, making it an appropriate choice for evaluating both baseline spectral features and the proposed SCD-based features. 

The SSL–SE-Res2Net50 system enables a direct comparison between handcrafted and SSL-derived two-dimensional feature representations when processed by the same high-capacity convolutional back-end. The SSL-only system provides a lightweight alternative that isolates the representational power of SSL embeddings with minimal back-end complexity. Finally, the embedding fusion-based system is designed to investigate the complementarity between convolutional and SSL representations. 

The first CM system (SE-Res2Net50) follows the standard front-end/back-end pipeline. Given an input waveform, frame-level acoustic features are extracted using a fixed feature extraction front-end. The resulting features are processed by an SE-Res2Net50 network adopted from \cite{Res2Net50_ASVspoof}, serving as a high-capacity feature encoder. The SE-Res2Net50 architecture comprises an initial convolutional stem composed of three $3\times3$ layers, followed by four stages of squeeze-and-excitation Res2Net bottleneck blocks (3–4–6–3). The SE-Res2Net blocks are followed by a global average pooling layer and a fully connected layer, producing a 256-dimensional embedding. ReLU activation is applied to all layers except the output layer, which uses softmax activation with two neurons representing bonafide and spoofed speech classes. 

The second CM system (SSL-SE-Res2Net50) replaces handcrafted acoustic features with SSL representations while keeping the back-end unchanged. A pre-trained Wav2Vec~2.0 base model \cite{wav2vec2} (trained on 960 hours of unlabelled Librispeech data \cite{Librispeech}) is used as the frozen SSL front-end. Wav2Vec~2.0 is selected as the SSL because the use of Librispeech is permitted under ASVspoof 5 evaluation rules since there is no speaker overlap with the ASVspoof 5 dataset. Frame-level features extracted from the 6th transformer encoder layer, which have a dimensionality of $200\times 768$, are used as the input to the SE-Res2Net50 network. The use of the 6th encoder layer output is motivated by previous studies \cite{layerwise_ssl, layerwise_ssl2} indicating that the intermediate SSL layers often encode richer spoofing-related cues than deeper layers, consequently improving detection performance.

The third, SSL-only CM system relies exclusively on the Wav2Vec~2.0 base model as a frozen frame-level feature extractor. The frame-level features are combined using a trainable layer-weighted averaging approach over all encoder layers and the initial convolutional layer following~\cite{stourbe24_asvspoof}. Temporal average pooling yields a $768$-dimensional utterance-level embedding, which is mapped to a compact 256-dimensional space using two linear projection layers before softmax classification. Only the layer weights, projection layers, and classifiers are trainable, which allows us to analyze the impact of SSL representations independently of convolutional modeling. 

The fourth and final system exploits the complementarity of convolutional and SSL representations through embedding-level fusion. Two parallel branches operate on the same input waveform. In the first branch, acoustic features are processed by a pre-trained and frozen SE-Res2Net50 model to obtain a $256$-dimensional SE-Res2Net50 embedding. In the second branch, the waveform is directly passed to the frozen Wav2Vec~2.0 model to obtain a $768$-dimensional SSL embedding computed using the same layer-weighted averaging and temporal average pooling strategy as in the SSL-only system. The dimensionality of the SSL embedding is reduced to 256 using a trainable linear layer to ensure a consistent embedding size with the SE-Res2Net50 embedding. Both embeddings are projected into a common 256-dimensional embedding space using a shared projection layer, which encourages alignment between the two modalities while reducing the number of trainable parameters. The final embedding is obtained using one of three fusion strategies:
\begin{itemize}[noitemsep, topsep=0pt]
\item \textbf{Element-wise weighted sum}: the two projected embeddings are combined using an element-wise weighted average approach via a learnable gating mechanism. A gating vector $\boldsymbol{\beta}\in \mathbb{R}^{256}$ is derived from the input embeddings to adaptively weight the contribution of each embedding per dimension. This strategy introduces only a modest number of additional parameters while providing greater flexibility than simple summation. 
\item \textbf{Element-wise summation (Add)}: a parameter-free fusion approach where the two embeddings are combined via simple element-wise addition. 
\item \textbf{Concatenation}: the two 256-dimensional embeddings are concatenated to form a 512-dimensional representation to increase the representational capacity while introducing only a small number of additional parameters compared to element-wise summation. 
\end{itemize}

The fused embedding is passed to a softmax classifier that outputs posterior probabilities for both classes. Across all systems, architectural and training configurations ensure fair, interpretable comparisons, so performance differences mainly reflect feature representations and fusion strategies rather than model capacity disparities. All systems are optimized using the Adam optimizer~\cite{adam_optimizer} with a learning rate of $1\times10^{-4}$ and early stopping based on development set EER.

\subsection{Performance Criteria} 
The performance of the SDD systems was evaluated using the EER and the minimum normalized detection cost function (minDCF) metrics \cite{ASVspoof5}. For the computation of minDCF, the prior probability of the spoof class and the associated costs were set to $\pi_\text{spf}=0.05$, $C_\text{miss}=1$, and $C_\text{fa}=10$, in accordance with the ASVspoof 5 challenge evaluation plan\footnote{\url{https://www.asvspoof.org/file/ASVspoof5___Evaluation_Plan_Phase2.pdf}}.

\section{Results}
\subsection{Results on the ASVspoof 2019 LA Dataset}
\label{sec:a19_results}

\begin{table}[!t]
\vspace{-3mm}
\centering
\caption{EER (\%) and minDCF on the ASVspoof 2019 LA development and evaluation sets for single-system CMs (Fig.~\ref{fig:cm_systems}): $(\textnormal{a})$ SE-Res2Net50 with different front-ends, $(\textnormal{b})$ SSL-SE-Res2Net50, and $(\textnormal{c})$ SSL-only with a lightweight back-end. For system $(\textnormal{a})$, best results across front-ends are in \textbf{bold}; best overall results are \underline{\textbf{bold and underlined}}. Training uses silence-trimmed training utterances; testing uses original utterances.}
\vspace{-3mm}
\label{tab:a19_results}
\tiny
\begin{tabular}{|l|l|cc|cc|}
\hline
\multirow{2}{*}{System} & \multirow{2}{*}{Front-end feature} &
\multicolumn{2}{c|}{Dev. Set} & \multicolumn{2}{c|}{Eval. Set} \\
\cline{3-6}
& & EER & minDCF & EER & minDCF \\
\hline

\multirow{7}{*}{(a) SE-Res2Net50}
& LFCC                       & 17.11 & 0.3924 & 22.88 & 0.4960 \\
& Mel-spectrogram            & 3.84  & 0.1010  & \textbf{12.07} & \textbf{0.2752} \\
& CQT                        & 6.16  & 0.1636  & 26.15 & 0.5395 \\
& STFT spectrogram           & 2.94  & 0.0770  & 35.01 & 0.5468 \\
\cline{2-6}
& SCD                        & \underline{\textbf{1.92}} & \underline{\textbf{0.0475}} & 22.88 & 0.5251 \\
& SCD$_a$                    & 24.21 & 0.5146 & 36.49 & 0.7451 \\
& SCD$_b$                    & 5.14  & 0.1261 & 19.78 & 0.4334 \\
\hline

(b) SSL-SE-Res2Net50
& SSL features               & 6.90  & 0.1690 & 11.62 & 0.3000 \\
\hline

(c) SSL-only
& SSL embeddings             & 5.68  & 0.1323 &
\underline{\textbf{8.28}} & \underline{\textbf{0.1978}} \\
\hline
\end{tabular}
\vspace{-3mm}
\end{table}

Table~\ref{tab:a19_results} reports EER and minDCF on the ASVspoof 2019 LA development and evaluation sets for the three single-system CMs, i.e., (a), (b), and (c) in Fig.~\ref{fig:cm_systems}. On the development set, the conventional SCD front-end yields the best SE-Res2Net50 performance, achieving 1.92\% EER and 0.0475 minDCF. The STFT follows with 2.94\% EER and 0.0770 minDCF. Among the baseline compact spectral features, the Mel-spectrogram performs best, followed by CQT and then LFCC. Within the SC family, the conventional SCD is best overall, while $\text{SCD}_b$ substantially outperforms $\text{SCD}_a$ and also improves over the LFCC and CQT baselines. This highlights the potential of SC representations in SDD tasks. For SSL-based systems, SSL-only outperforms SSL-SE-Res2Net50 on both metrics, indicating that a lightweight back-end can be sufficient when using SSL representations.

On the evaluation set, the SSL-only CM achieves the best single-system performance (8.28\% EER, 0.1978 minDCF). Among the SE-Res2Net50 acoustic front-ends, Mel-spectrogram yields the lowest EER (12.07\%) and minDCF (0.2752), followed by $\text{SCD}_b$ and conventional SCD. A discrepancy is observed between the development and evaluation set results, where SCD performs best on the development set but degrades relative to other features on the evaluation set. This can be attributed to differences in representation where SCD captures global $(f,\alpha)$ cyclostationary structure through time averaging, which is effective under matched conditions. Representations preserving temporal dynamics (e.g., $\text{SCD}_b$), in turn, generalize better to unseen attacks. In contrast, STFT spectrogram features degrade sharply (35.01\% EER), indicating their limited generalization. Overall, these results suggest a generalization advantage for SSL embeddings. Interestingly, the SSL-only CM outperforms SSL-SE-Res2Net50 on the evaluation set (8.28\% vs. 11.62\% EER), corresponding to a relative EER reduction of approximately 29\%. Consequently, SSL-only results are used as the primary SSL baseline in the remainder of this work.

\begin{figure}[!h]
\vspace{-3mm}
\centerline{\includegraphics[scale=0.20]
{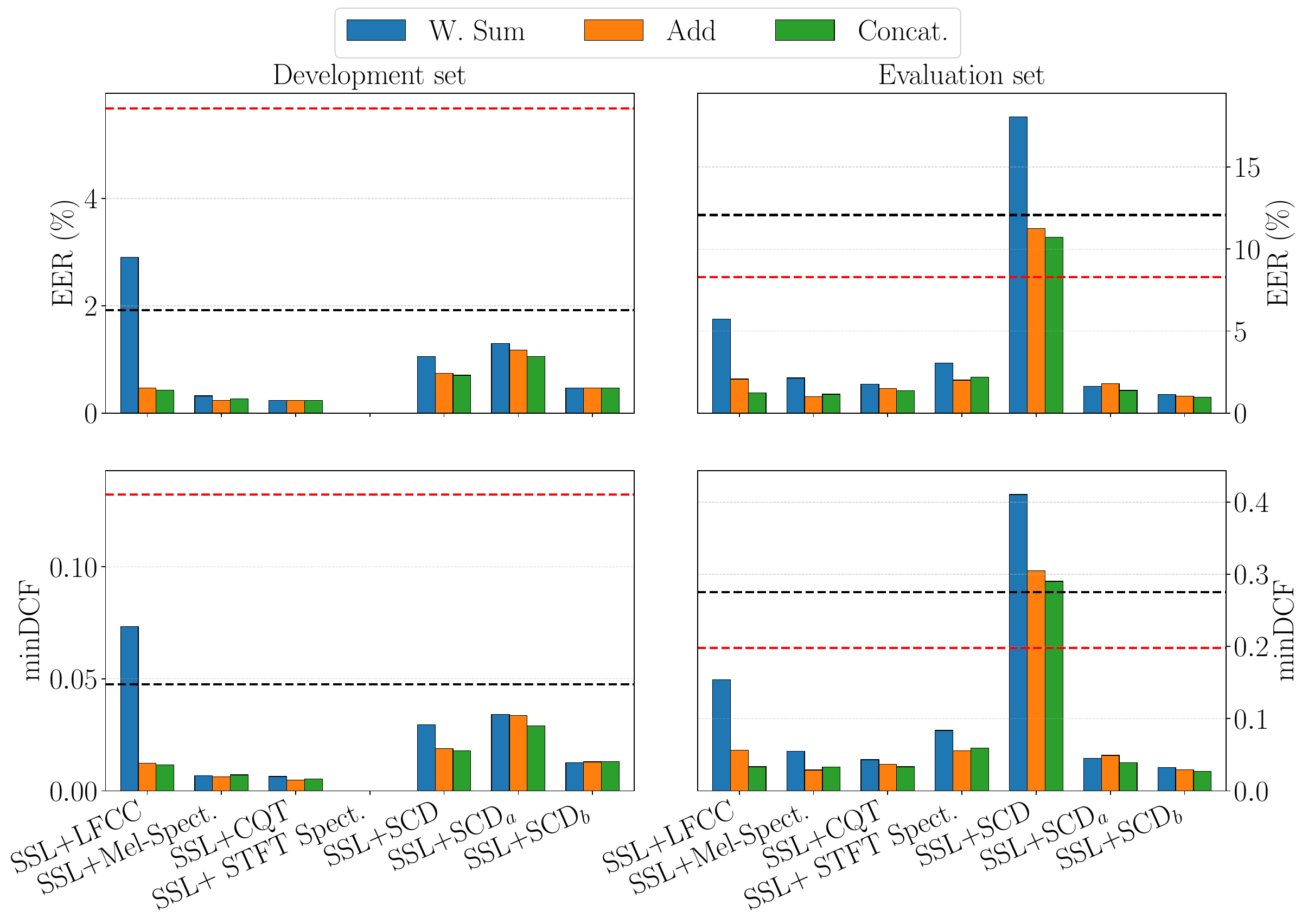}}
\vspace{-3mm}
\caption{ASVspoof 2019 LA embedding-fusion results (Fig.~\ref{fig:cm_systems}$(\textnormal{d})$) on the development (left) and evaluation (right) sets. Top and bottom panels show EER and minDCF, respectively. Dashed lines indicate the best single-system baselines: SCD (black) and SSL-only (red) for the development set, and Mel-spectrogram (black) and SSL-only (red) for the evaluation set.}
\label{fig:ssl_fusion_results_a19}
\vspace{-3mm}
\end{figure}

Beyond the single-system baselines, we evaluate the embedding fusion CM, with the results summarized in Fig.~\ref{fig:ssl_fusion_results_a19} for the three fusion strategies. Across front-ends, concatenation consistently provides the strongest and most stable evaluation set performance, indicating that increased representational capacity at the fusion stage is beneficial. On the evaluation set, concatenation achieves EERs of 1.23\%, 1.55\%, 1.36\%, 2.20\%, 1.38\%, and 0.98\% when combined with LFCC, Mel-spectrogram, CQT, STFT spectrogram, $\text{SCD}_a$, and $\text{SCD}_b$, respectively. These improvements over the single-system results (Table~\ref{tab:a19_results}) indicate strong complementarity between SE-Res2Net50 and SSL embeddings. Among all evaluated configurations, concatenation-based fusion with $\text{SCD}_b$ achieves the lowest evaluation set EER and minDCF. 

In contrast, weighted-sum fusion exhibits larger variability and generally inferior evaluation set performance, with element-wise addition producing intermediate results. Relative to the strongest single-system baselines (horizontal reference lines), all fusion configurations outperform both the SSL-only CM and the best SE-Res2Net50 acoustic baseline except for the conventional SCD. The best fused system (SSL+$\text{SCD}_b$ with concatenation) achieves 0.98\% EER on the evaluation set, corresponding to EER reductions of approximately 95\% and 88\% relative to SE-Res2Net50+$\text{SCD}_b$ (19.78\%) and the SSL-only CM (8.28\%), respectively. Based on these findings, only concatenation-based fusion results are reported in the remainder of the paper. 

To further analyze the proposed SC-based features and their interaction with SSL embeddings, Table~\ref{tab:a19_attack_wise_results} reports attack-wise EERs on the ASVspoof 2019 LA evaluation set for the SSL-only system and the selected concatenation-based fusion systems: SSL+LFCC, SSL+$\text{SCD}_a$, and SSL+$\text{SCD}_b$. The results indicate that fusing $\text{SCD}_b$ embeddings with SSL embeddings consistently improves spoofing detection across a wide range of attacks and achieves the lowest pooled EER (0.98\%), compared with 8.28\% for SSL-only, 1.23\% for SSL+LFCC, and 1.38\% for SSL+$\text{SCD}_a$. Compared with LFCC, the proposed SC-based features also achieve lower EERs for most attacks, suggesting that they provide more discriminative complementary information to SSL embeddings. Overall, these results indicate that the gains from combining SC-based and SSL embeddings generalize across diverse spoofing attacks rather than being limited to specific attack conditions.

\begin{table}[t]
\centering
\caption{Attack-wise EERs (\%) on the ASVspoof 2019 LA evaluation set for the SSL-only system (Fig.~\ref{fig:cm_systems}$(\textnormal{c})$) and the selected concatenation-based fusion systems (Fig.~\ref{fig:cm_systems}$(\textnormal{d})$): SSL+LFCC, SSL+$\text{SCD}_a$, and SSL+$\text{SCD}_b$. The lowest EER for each attack is shown in \textbf{bold}; the final row reports pooled EERs.}
\vspace{-2mm}
\label{tab:a19_attack_wise_results}
\tiny
\begin{tabular}{|l|c|c|c|c|}
\hline
Attack & \multicolumn{4}{c|}{System} \\ \cline{2-5}
Type & (c) SSL-only & (d) SSL+LFCC & (d) SSL+$\text{SCD}_a$ & (d) SSL+$\text{SCD}_b$  \\
\hline
A07    & 8.50 & 0.26 & 0.30 & \textbf{0.16} \\
A08    & 3.37 & 2.48 & 2.07 & \textbf{1.34} \\
A09    & 3.09 & 0.09 & 0.12 & \textbf{0.05} \\
A10    & 29.22 & 1.44 & 1.14 & \textbf{0.46} \\
A11    & 6.61 & 0.36 & \textbf{0.26} & 0.32 \\
A12    & 9.66 & 0.34 & 0.32 & \textbf{0.20} \\
A13    & 0.26 & 0.08 & \textbf{0.01} & 0.04 \\
A14    & 2.09 & 0.95 & 0.24 &\textbf{0.12} \\
A15    & 5.59 & 1.28 & 0.83 &\textbf{0.34} \\
A16    & 6.30 & 0.50 & 0.54 &\textbf{0.38} \\
A17    & \textbf{2.25} & 2.27 & 3.15 &  2.62 \\
A18    & \textbf{0.58} & 1.28 & 1.62 & 0.97 \\
A19    & 8.36 & 1.40 & 1.60 &\textbf{0.75} \\
\hline
Pooled & 8.28 & 1.23 & 1.38 &\textbf{0.98} \\
\hline
\end{tabular}
\end{table}

\subsection{Results on the ASVspoof 2021 DF Dataset}

\begin{table*}[!t]
\vspace{-3mm}
\centering
\caption{EER (\%) by vocoder category on the ASVspoof 2021 DF dataset. Results are reported for the systems in Fig.~\ref{fig:cm_systems}: $(\textnormal{a})$ SE-Res2Net50 with different acoustic front-ends, $(\textnormal{c})$ SSL-only, and $(\textnormal{d})$ concatenation-based embedding fusion. Columns represent Neural AR, Neural non-AR, DSP-based, waveform concatenation, and unknown vocoders; the final columns report pooled EER and minDCF. The highest EER for each system is shaded, and the best result in each column is shown in \textbf{bold}. Training uses silence-trimmed utterances, whereas testing uses the original utterances.}
\label{tab:a21_df_results}
\vspace{-2mm}
\tiny
\begin{tabular}{|l|l|ccccc|cc|}
\hline
\multirow{2}{*}{System} & \multirow{2}{*}{Front-end / Fusion} &
\multicolumn{5}{c|}{Vocoder Type (EERs in \%)} & \multicolumn{2}{c|}{Pooled} \\
\cline{3-9}
 &  & Neural AR & Neural non-AR & DSP-based & Wav.~Concat. & Unknown & EER & minDCF \\
\hline

\multirow{7}{*}{(a) SE-Res2Net50}
& LFCC              & 30.89 & 31.34 & 27.77 & \rowmax{34.50} & 24.68 & 29.39 & 0.6871 \\
& Mel-spectrogram   & 36.52 & \rowmax{37.07} & 23.49 & 18.07 & 24.95 & 30.17 & 0.6923 \\
& CQT               & \rowmax{42.25} & 37.95 & 18.66 & 17.94 & 21.29 & 30.06 & 0.6827 \\
& STFT spectrogram  & \rowmax{49.71} & 41.85 & 33.61 & 24.91 & 32.23 & 39.33 & 0.8839 \\
\cline{2-9}
& SCD               & \rowmax{41.58} & 40.35 & 37.75 & 27.29 & 29.81 & 38.75 & 0.9995 \\
& SCD$_a$           & 47.05 & 47.59 & 29.20 & \rowmax{47.85} & 24.93 & 37.87 & 0.8649 \\
& SCD$_b$           & \rowmax{31.85} & 28.61 & 13.97 & 19.21 & \textbf{9.99} & 22.73 & 0.5596 \\
\hline

(c) SSL-only
& SSL embeddings
& \textbf{\rowmax{13.90}} & \textbf{10.99} & \textbf{4.55} & \textbf{8.03} & 13.81
& \textbf{9.77} & \textbf{0.2363} \\
\hline

\multirow{7}{*}{(d) Embedding fusion}
& SSL + LFCC        & \rowmax{14.89} & 12.79 & 9.48 & 7.64 & 12.73 & 12.12 & 0.3159 \\
& SSL + Mel-spectrogram
                    & \rowmax{17.86} & 17.85 & 11.37 & 11.07 & 13.26 & 15.38 & 0.4201 \\
& SSL + CQT         & \rowmax{18.74} & 15.83 & 7.25 & 14.50 & 13.08 & 13.58 & 0.3614 \\
& SSL + STFT spectrogram
                    & \rowmax{25.60} & 22.20 & 8.88 & 12.03 & 10.27 & 17.77 & 0.4364 \\
\cline{2-9}
& SSL + SCD         & \rowmax{36.33} & 33.16 & 26.07 & 23.48 & 15.79 & 29.88 & 0.7078 \\
& SSL + SCD$_a$     & \rowmax{15.21} & 14.29 & 7.97 & 7.88 & 13.86 & 12.24 & 0.3220 \\
& SSL + SCD$_b$     & \rowmax{17.61} & 16.10 & 9.89 & 14.16 & 15.25 & 14.39 & 0.3900 \\
\hline
\end{tabular}
\vspace{-3mm}
\end{table*}

Table~\ref{tab:a21_df_results} reports ASVspoof 2021 DF results both per vocoder type and in pooled form. For SE-Res2Net50, $\text{SCD}_b$ achieves the best overall performance, with a pooled EER of 22.73\% and minDCF of 0.5596. This corresponds to improvements of $\sim$23\% (EER) and $\sim$19\% (minDCF) relative to the second-best SE-Res2Net50 model based on LFCC features.

Across the vocoder categories, $\text{SCD}_b$ consistently outperforms other spectral representations for attacks generated by neural non-autoregressive, DSP-based, and unknown vocoders. Although LFCC yields a slightly lower EER for neural autoregressive vocoders, the performance difference relative to  $\text{SCD}_b$ is modest (30.89\% vs.\ 31.85\%), indicating comparable robustness in this category.

The SSL-only CM achieves uniformly low EERs across most vocoder types, especially for DSP-based and waveform concatenation attacks. It yields a pooled EER of 9.77\% and a minDCF of 0.2363, representing the best overall performance among all systems. In contrast, $\text{SCD}_b$ with the SE-Res2Net50 classifier yields considerably lower EERs for unknown vocoder attacks compared with the SSL-only CM. This indicates that SSL embeddings capture vocoder-agnostic cues that generalize to diverse unseen attacks, whereas $\text{SCD}_b$ is particularly effective for the unknown vocoder category.

Consistent trends are observed across systems with respect to vocoder difficulty. Neural vocoders are generally the most challenging, while waveform concatenation attacks are easier for conventional acoustic features. The DSP-based vocoders, in turn, are best detected using $\text{SCD}_b$, while neural non-autoregressive attacks are of intermediate difficulty.

Unlike ASVspoof 2019, embedding fusion does not provide systematic improvements over the SSL-only CM. In most cases, the pooled EER and minDCF increase relative to the SSL-only system, regardless of the acoustic front-end. While some fused systems (e.g., SSL+LFCC and SSL+$\text{SCD}_a$) retain competitive performance, none surpass the SSL-only CM.

An interesting observation is that, although the standalone $\mathrm{SCD}_a$ representation performs worse than $\mathrm{SCD}_b$, the SSL+$\mathrm{SCD}_a$ fusion system achieves the best performance among the SC-based fusion approaches. This suggests that $\mathrm{SCD}_a$ captures complementary cyclic modulation information that is not sufficiently discriminative on its own---but becomes effective when combined with SSL embeddings. Since $\mathrm{SCD}_a$ aggregates spectral correlation information across the spectral frequency axis, it emphasizes temporal modulation characteristics while discarding fine spectral localization. Although this aggregation reduces standalone discriminability, it may provide complementary information to SSL embeddings, which already capture strong spectral structure. In contrast, $\mathrm{SCD}_b$ preserves more spectral detail and may therefore introduce greater redundancy during fusion.

Nevertheless, embedding fusion substantially improves performance relative to SE-Res2Net50-based CMs alone, indicating that SSL embeddings remain the dominant source of discriminative information.

\subsection{Results on the ASVspoof5 Dataset}
\begin{table}[!t]
\vspace{-3mm}
\centering
\caption{EER (\%) and minDCF on the ASVspoof 5 development set.
Results are reported for the systems shown in Fig.~\ref{fig:cm_systems}: $(\textnormal{a})$ SE-Res2Net50 with different acoustic front-ends, $(\textnormal{c})$ an SSL-only CM, and $(\textnormal{d})$ concatenation-based embedding-fusion CMs combining SSL and SE-Res2Net50 embeddings.}
\vspace{-2mm}
\label{tab:asvspoof5_dev_results}
\tiny
\begin{tabular}{|l|l|cc|}
\hline
System & Front-end / Fusion & EER & minDCF \\
\hline

\multirow{7}{*}{(a) SE-Res2Net50}
& LFCC             & 17.00 & 0.3101 \\
& Mel-spectrogram  & 17.47 & 0.2675 \\
& CQT              & 27.70 & 0.4841 \\
& STFT spectrogram & 9.22  & 0.1296 \\
\cline{2-4}
& SCD              & 42.52 & 0.8591 \\
& SCD$_a$          & 29.26 & 0.6530 \\
& SCD$_b$          & 23.17 & 0.4752 \\
\hline

(c) SSL-only
& SSL embeddings   & \textbf{5.63} & \textbf{0.1358} \\
\hline

\multirow{7}{*}{(d) Embedding fusion}
& SSL + LFCC             & 15.89 & 0.2890 \\
& SSL + Mel-spectrogram  & 16.71 & 0.2338 \\
& SSL + CQT              & 15.64 & 0.3295 \\
& SSL + STFT spectrogram & 8.20  & 0.1302 \\
\cline{2-4}
& SSL + SCD              & 10.42 & 0.2544 \\
& SSL + SCD$_a$          & 11.51 & 0.2908 \\
& SSL + SCD$_b$          & 14.20 & 0.2741 \\
\hline
\end{tabular}
\vspace{-3mm}
\end{table}

Our final evaluation results on the development portion of ASVspoof 5 are summarized in Table~\ref{tab:asvspoof5_dev_results}. The error rates are considerably higher than on ASVspoof 2019, as anticipated. Among the SE-Res2Net50-based systems, the STFT spectrogram yields the best performance, achieving EER 9.22\% and minDCF 0.1296. LFCC and Mel-spectrogram provide comparable EERs, with Mel-spectrogram yielding a lower minDCF. Within the SC family, $\text{SCD}_b$ outperforms both the conventional SCD and $\text{SCD}_a$. It also outperforms CQT in both metrics, indicating its relative robustness.

Comparing the single-system baselines, SSL-only achieves the best performance, with an EER of 5.63\% and a minDCF of 0.1358, corresponding to a 39\% reduction in EER relative to the best SE-Res2Net50 configuration (STFT spectrogram). This again highlights the strong generalization capability of pretrained SSL representations.

The fusion approach consistently improves over the corresponding SE-Res2Net50 baselines for most front-ends. In particular, fusion with STFT spectrogram and conventional SCD embeddings yields the strongest fusion performance, achieving EERs of 8.20\% and 10.42\%, respectively. Despite these gains, all fusion variants remain inferior to SSL-only, suggesting that SSL embeddings already capture highly discriminative spoofing cues on the development set.

Proceeding to the evaluation set, the attack-wise results, together with pooled EER and minDCF, are reported in Table~\ref{tab:asvspoof5_eval_results} for selected systems based on development set performance: SSL-only as the primary reference; SSL+CQT as the best-performing fusion system with a compact spectral representation; SSL+STFT spectrogram as the best-performing fusion system based on STFT spectrograms, and SSL+SCD as the best-performing SC fusion system. For reference purposes, we also include the official ASVspoof~5 baselines (B01: RawNet2 and B02: AASIST) \cite{ASVspoof5}. 

In contrast to the development set results, the embedding fusion approach with conventional SCD representation yields the best evaluation set performance, achieving a pooled EER of 14.80\% and a minDCF of 0.3423, outperforming the SSL-only CM (15.81\% EER, 0.3631 minDCF). The attack-wise analysis reveals that SSL+SCD improves performance on 11 out of 16 attacks (A17--A20, A22, A25--A27, and A30--A32), including six of the seven adversarial attacks (A18, A20, A27, and A30--A32). This indicates that conventional SCD captures spoofing cues that are particularly relevant for adversarial attack scenarios and that complement the information encoded in SSL embeddings. In contrast, although the SSL+STFT spectrogram configuration performed well on the development set, its performance degrades substantially on the evaluation set. This shows the limited generalization capability of STFT spectrograms on unseen and diverse attacks.

\begin{table*}
\vspace{-3mm}
\begin{center}
    \caption{Attack-wise and pooled EER (\%) and minDCF on the ASVspoof 5 evaluation set for SSL-only and selected concatenation-based fusion systems. Pooled metrics are computed from the scores pooled across all attacks. The lowest values per attack/metric are in \textbf{bold}. Challenge baselines B01 (RawNet2) and B02 (AASIST) are from \cite{ASVspoof5}. The last three rows report relative change (\%) w.r.t.\ SSL-only; negative values (green) indicate improvements, while positive values (red) denote degradation.}
    \vspace{-3mm}
    \label{tab:asvspoof5_eval_results}
    \tiny
    \resizebox{\textwidth}{!}{
    \begin{tabular}{|l|cccccccccccccccc|cc|}\hline
      System & A17 & A18 & A19 & A20 & A21 & A22 & A23 & A24 & A25 & A26 & A27 & A28 & A29 & A30 & A31 & A32 & EER & minDCF \\ \hline
      SSL-only (c) & 5.42  & 11.60  & 8.51  & 9.82 & 8.99 & \textbf{10.54} & \textbf{19.61}& \textbf{27.56}& 5.17 & 7.13 & 15.44  & 25.00   &5.49 & 24.46 & 37.38 & 12.33 & 15.81 & 0.3631\\ \hline
     SSL + CQT (d)  & 5.93 & 29.67 & 20.88  & 29.27 & \textbf{4.78}  & 15.25 & 31.92 & 57.76 & 14.46 & 19.77 & 41.16  &  \textbf{21.01} & \textbf{3.03} &  42.41 & 38.32 & 33.45 & 27.54 & 0.6915 \\
     SSL + STFT Spect. (d) & 33.10 & 36.39 & 46.22 & 50.82 & 33.96 & 36.11 & 36.24 & 38.74 & 37.84  & 33.74  &  47.60 &  35.91 & 32.24 & 47.49 & 47.44 & 50.95 & 39.56 & 0.9313 \\
     SSL + SCD (d) & \textbf{3.78} & \textbf{7.73} & \textbf{4.68} & \textbf{5.61} & 9.02 & \textbf{10.54} & 21.43 & 60.01 & \textbf{3.93} & \textbf{4.79} & \textbf{6.95}  & 31.46  & 5.16 & \textbf{14.57} & \textbf{32.26} & \textbf{5.67} & \textbf{14.80} & \textbf{0.3423} \\ \hline
     \hline
     ASVspoof 5 B01 \cite{ASVspoof5} & 22.58 & 57.64 & 63.75 & 46.95 & 25.67 & 24.50 & 30.35 & 23.61 & 29.78 & 41.95 & 38.85 & 39.43 & 17.54 & 42.26 & 33.22 & 29.78 & 36.04 & 0.8266 \\
     ASVspoof 5 B02 \cite{ASVspoof5} & 16.44 & 50.02 & 59.99 & 33.70 & 17.05 & 17.63 & 32.78 & 13.35 & 21.01 & 31.35 & 27.60 & 32.10 & 8.93 & 42.85 & 27.48 & 19.50 & 29.12 & 0.7106 \\ \hline 
\multicolumn{19}{|c|}{\textit{\textbf{Relative change w.r.t. SSL-only} (\%)}} \\ \hline

$\Delta$ SSL + CQT
& \relpos{+9.4}  & \relpos{+155.7} & \relpos{+145.4} & \relpos{+198.1}
& \relneg{-46.8} & \relpos{+44.7}  & \relpos{+62.8}  & \relpos{+109.6}
& \relpos{+179.7}& \relpos{+177.3} & \relpos{+166.6} & \relneg{-16.0}
& \relneg{-44.8} & \relpos{+73.3}  & \relpos{+2.5}   & \relpos{+171.3}
& \relpos{+74.2} & \relpos{+90.4} \\

$\Delta$ SSL + STFT Spect.
& \relpos{+510.7}& \relpos{+213.7} & \relpos{+443.1} & \relpos{+417.5}
& \relpos{+277.8}& \relpos{+242.6} & \relpos{+84.8}  & \relpos{+40.6}
& \relpos{+631.9}& \relpos{+373.2} & \relpos{+208.3} & \relpos{+43.6}
& \relpos{+487.2}& \relpos{+94.1}  & \relpos{+26.9}  & \relpos{+313.2}
& \relpos{+150.2}& \relpos{+156.5} \\

$\Delta$ SSL + SCD
& \relneg{-30.2} & \relneg{-33.4} & \relneg{-45.0} & \relneg{-42.9}
& \relpos{+0.3}  & \relpos{+0.0}  & \relpos{+9.3}  & \relpos{+117.7}
& \relneg{-24.0} & \relneg{-32.8} & \relneg{-55.0} & \relpos{+25.8}
& \relneg{-6.0}  & \relneg{-40.4} & \relneg{-13.7} & \relneg{-54.0}
& \relneg{-6.4}  & \relneg{-5.7} \\
\hline

    \end{tabular}
    }
  \end{center}
  \vspace{-3mm}
\end{table*}

To place the proposed framework in a broader context, state-of-the-art systems on ASVspoof5 dataset yield evaluation set EERs ranging from approximately 10\% to 22\% using large-scale SSL-based, Mamba-based, and proprietary architectures as summarized in Table~2 of \cite{speech_df_arena}. For reference, the best-performing system investigated in this work, obtained by concatenating SSL and $\mathrm{SCD}$ embeddings, achieved an EER of 14.80\%. Although direct numerical comparisons should be interpreted with caution due to differences in training protocols, model capacity, data augmentation, and implementation details, these results indicate that the proposed framework performs competitively with several recent open-source systems.

Furthermore, we also evaluated the same concatenation-based embedding fusion strategy using the AASIST \cite{aasist} and AASIST3 \cite{aasist3} architectures. While these implementations produced considerably higher absolute EERs than the SSL-SE-Res2Net50 framework, the $\mathrm{SCD}_b$ emmbeddings consistently provided complementary information, reducing the EER from 32.50\% to 30.15\% for AASIST and from 38.71\% to 34.52\% for AASIST3. This observation further suggests that the benefit of higher-order cyclostationary information is not tied to a particular backbone architecture, but rather provides complementary information beyond modern learned representations.

Motivated by the strong evaluation-set generalization of the concatenation-based SSL+SCD framework, we further investigated whether more sophisticated interaction mechanisms between SSL and cyclostationary representations could yield additional improvements. Inspired by multimodal representation-learning approaches based on cross-modal attention and selective feature routing \cite{cross_attention, windowed_attention}, we implemented and evaluated two alternative fusion strategies within the proposed utterance-level SSL/SCD embedding framework: \emph{cross-attentive fusion} and \emph{windowed binary gating}. In the cross-attentive configuration, SSL and SCD embeddings mutually attend to each other using bidirectional multi-head attention prior to classification. More specifically, the two 256-dimensional embeddings are first projected into a shared latent space and processed using a bottlenecked cross-attention module with four attention heads and an intermediate attention dimensionality of 64. Residual connections and layer normalization are then applied before a lightweight feed-forward fusion block produces the final fused representation. In the windowed binary gating approach, the fused representation is obtained through learned element-wise selection between SSL and SCD embeddings. The 256-dimensional embeddings are partitioned into non-overlapping windows of size 16, and a gating network generates soft binary masks that determine which embedding source dominates each local region. All implementations were trained under the same optimization settings as the concatenation-based baseline.

Table~\ref{tab:additional_fusion_results} summarizes the pooled EERs obtained with these alternative fusion strategies. Although cross-attentive fusion achieved lower development-set EERs for some configurations, its evaluation-set performance was consistently inferior to concatenation-based fusion, indicating weaker robustness to unseen attacks. The windowed binary gating approach, in turn, produced substantially higher EERs on both development and evaluation sets. Overall, these findings suggest that the simpler concatenation-based strategy generalizes better than the more constrained attention- or gating-based alternatives. One possible explanation is that the fusion is performed at the utterance-embedding level rather than at the frame/token level, thereby limiting the temporal or local structural information available for attention-based interaction modeling. In addition, cross-attention introduces substantially higher computational complexity and training overhead due to bidirectional multi-head attention, while providing no consistent evaluation-set benefit in the present spoofing detection setting.

\begin{table}[!t]
    \centering
     \caption{Pooled EERs (\%) on the ASVspoof~5 development and evaluation sets for different SSL/SCD embedding-fusion strategies, including concatenation-based (Concat.), cross-attentive (Cross-attn.), and windowed binary gating (Win. gating) fusion approaches.}
     \vspace{-0.25cm}
    \label{tab:additional_fusion_results}
    \scriptsize
    \begin{tabular}{|l|l|cc|}
    \hline
    Fusion Type & System & Dev. EER & Eval. EER \\
    \hline
    \multirow{3}{*}{Concat.}
    & SSL + SCD       & 10.42 & \textbf{14.80} \\
    & SSL + SCD$_a$   & 11.51 & \textbf{22.13} \\
    & SSL + SCD$_b$   & 14.20 & \textbf{23.51} \\
\hline
    \multirow{3}{*}{Cross-attn.}
    & SSL + SCD       & \textbf{7.46} & 16.65 \\
    & SSL + SCD$_a$   & \textbf{10.63} & 22.36 \\
    & SSL + SCD$_b$   & \textbf{11.82} & 26.28 \\
\hline
    \multirow{3}{*}{Win. gating}
    & SSL + SCD       & 12.07 & 22.24 \\
    & SSL + SCD$_a$   & 28.62 & 37.53 \\
    & SSL + SCD$_b$   & 23.71 & 35.43 \\
\hline
\end{tabular}
\vspace{-0.25cm}
\end{table}

The observed improvements obtained through various embedding fusion suggest that SCD representations and SSL embeddings encode complementary speech aspects. While SSL models learn highly effective speech representations through large-scale pretraining, they are optimized using data-driven objectives rather than explicit signal-processing principles. In contrast, SCD explicitly characterizes higher-order spectral dependencies arising from cyclostationary structure. Consequently, the two representations provide complementary rather than competing descriptions of the speech signal. This signal-processing perspective explains why combining SCD and SSL representations can improve performance across multiple datasets and backbone architectures despite the simplicity of the concatenation-based fusion strategy.

\begin{figure}[!h]
\vspace{-3mm}
\centerline{\includegraphics[scale=0.18]{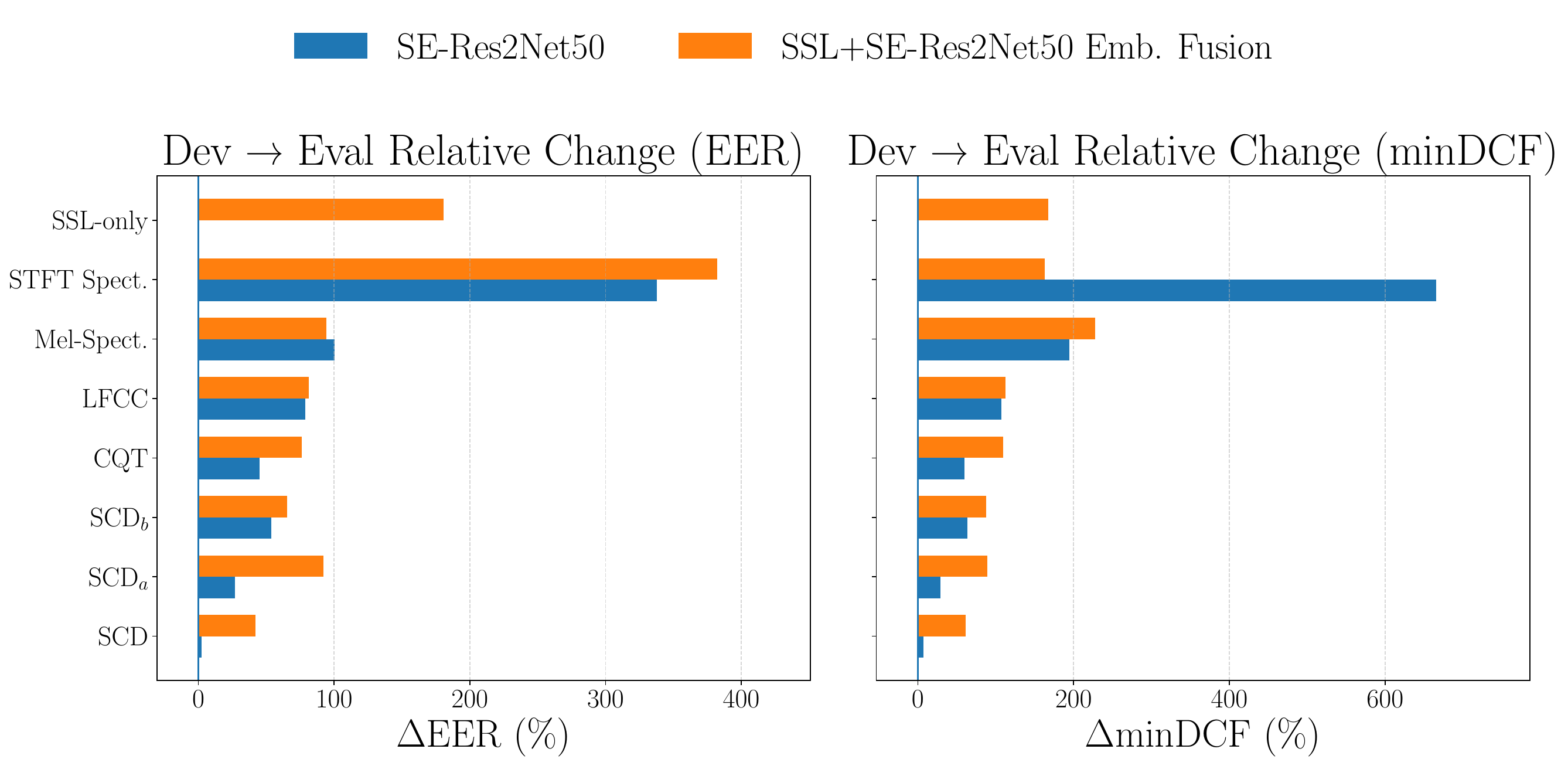}}
\vspace{-3mm}
\caption{Relative change (\%) in EER (top) and minDCF (bottom) when moving from the development set to the evaluation set on ASVspoof 5, for different acoustic feature representations using SE-Res2Net50 and concatenation-based embedding-fusion CMs. Smaller relative changes indicate better generalization under unseen and diverse spoofing attacks.}
\label{fig:asvspoof5_generalization_metric}
\vspace{-3mm}
\end{figure}

To further assess generalization, Fig.~\ref{fig:asvspoof5_generalization_metric} shows the relative degradation from the development set to the evaluation set. As is evident, the SCD features, particularly the conventional SCD, exhibit consistently smaller relative degradation than the other acoustic features for both SE-Res2Net50-based and fusion-based systems. This reinforces that SCD features offer improved robustness to unseen and diverse spoofing attacks.

To analyze the computational overhead introduced by the proposed and best-performing concatenation-based embedding-fusion framework, Table~\ref{tab:inference_time_asvspoof5} reports inference-time statistics for the SSL-only and SSL+$\text{SCD}$ systems on the ASVspoof~5 evaluation set. The measurements were obtained using 10000 randomly selected utterances and averaged over three independent runs. The experiments were conducted using batch size 1 on a system equipped with an NVIDIA RTX 4000 Ada GPU, an Intel i9-14900K CPU, and 128\,GB RAM. The reported timings correspond only to the model forward-pass inference stage. Although SSL+$\text{SCD}$ approximately doubles the mean inference time relative to SSL-only (24.68\,ms versus 11.20\,ms per utterance), both systems remain substantially faster than real time, corresponding to approximately 286$\times$ and 625$\times$ real-time processing speed, respectively. The reported measurements exclude offline SCD embedding extraction. Therefore, despite the additional computational cost introduced by embedding fusion, the inference latency remains very low. Considering that the SSL+SCD system achieves a relative pooled EER reduction of 6.4\% compared with the SSL-only baseline, while also providing attack-wise relative EER improvements ranging from 6.0\% up to 55.0\% for several attacks (Table~\ref{tab:asvspoof5_eval_results}), particularly adversarial attacks, the additional computational overhead can be deemed acceptable for practical applications.

\begin{table}[!t]
\centering
\caption{Inference-time comparison of SSL-only and SSL+$\text{SCD}$ fusion systems on the ASVspoof5 evaluation set.}
\label{tab:inference_time_asvspoof5}
\vspace{-2mm}
\scriptsize
\begin{tabular}{|l|c|c|c|}
\hline
System & Mean Inf. Time & Std. Dev. & RTF \\
\hline
SSL-only & 11.20 ms/utt & 2.13 ms & 0.0016 \\
SSL + $\text{SCD}$ & 24.68 ms/utt & 6.32 ms & 0.0035 \\
\hline
\end{tabular}
\vspace{-3mm}
\end{table}

\subsection{Cross-Dataset Evaluation}
To further evaluate the robustness and generalization capability of the proposed SC-based features, we conducted cross-dataset experiments using the ASVspoof 2019 LA (A19) and ASVspoof~5 (A5) datasets. In these experiments, systems were trained on one dataset and evaluated on the other, introducing substantial mismatch in spoofing methods, acoustic environments, recording conditions, and channel characteristics. Table~\ref{tab:cross_dataset_results} summarizes the pooled EER results for both the cross-dataset evaluation experiments and the corresponding matched-condition experiments, in which the systems were trained and evaluated on the same dataset, to facilitate direct comparison.

\begin{table}[t]
\centering
\caption{Cross-dataset evaluation results in terms of pooled EER (\%). Cross denotes training and testing on different datasets, while
Matched denotes training and testing on the same dataset.
A5 and A19 denote the ASVspoof 5 and ASVspoof 2019 LA datasets, respectively.}
\label{tab:cross_dataset_results}
\vspace{-0.25cm}
\scriptsize
\setlength{\tabcolsep}{4pt}
\begin{tabular}{|c|l|cc|cc|}
\hline
\multirow{2}{*}{Train $\rightarrow$ Test} &
\multirow{2}{*}{System} &
\multicolumn{2}{c|}{Dev. EER} &
\multicolumn{2}{c|}{Eval. EER} \\
\cline{3-6}
& & Cross & Matched & Cross & Matched \\
\hline
\multirow{8}{*}{A19 $\rightarrow$ A5}
& SSL-only            & 9.61 & 5.63  & 22.51 & 15.81 \\
& SSL+LFCC            & 3.74 & 15.89 & 11.56 & 28.80 \\
& SSL+Mel-Spect.      & 3.86 & 16.71 & 20.94 & 32.45 \\
& SSL+CQT             & 11.32 & 15.64 & 22.53 & 27.54 \\
& SSL+STFT            & 4.00 & 8.20  & 38.38 & 39.56 \\
& SSL+SCD             & 19.93 & 10.42 & 29.52 & 14.80 \\
& SSL+$\text{SCD}_a$  & 7.04 & 11.51 & 17.07 & 22.13 \\
& SSL+$\text{SCD}_b$  & 7.75 & 14.20 & 18.94 & 23.51 \\
\hline
\multirow{8}{*}{A5 $\rightarrow$ A19}
& SSL-only            & 28.21 & 5.68 & 32.92 & 8.28 \\
& SSL+LFCC            & 0.81 & 0.43 & 4.76 & 1.23 \\
& SSL+Mel-Spect.      & 0.39 & 0.26 & 2.35 & 1.15 \\
& SSL+CQT             & 0.42 & 0.23 & 1.73 & 1.36 \\
& SSL+STFT            & 0.02 & 0.00 & 3.24 & 2.20 \\
& SSL+SCD             & 1.88 & 0.70 & 22.20 & 10.72 \\
& SSL+$\text{SCD}_a$  & 1.29 & 1.05 & 3.51 & 1.38 \\
& SSL+$\text{SCD}_b$  & 0.61 & 0.47 & 1.71 & 0.97 \\
\hline
\end{tabular}
\vspace{-0.25cm}
\end{table}

Overall, the results indicate that incorporating handcrafted acoustic representations through embedding fusion substantially improves cross-dataset spoofing detection compared with the SSL-only baseline. In the A5$\rightarrow$A19 setting, all fusion systems significantly outperform SSL-only, whose evaluation EER reaches 32.92\%. SSL+$\text{SCD}_b$ achieves the best performance with an evaluation EER of 1.71\%, followed by SSL+CQT and SSL+Mel-spectrogram. Similarly, in the A19$\rightarrow$A5 setting, SSL+$\text{SCD}_a$ and SSL+$\text{SCD}_b$ significantly outperform SSL-only, reducing the evaluation EER from 22.51\% to 17.07\% and 18.94\%, respectively. In contrast, although SSL+STFT achieves competitive performance under matched conditions, its cross-dataset performance remains relatively weak, particularly in the A19$\rightarrow$A5 configuration.

An interesting asymmetric transfer behavior is also observed between the two cross-dataset configurations. In the A19$\rightarrow$A5 setting, several systems, including SSL+LFCC, SSL+$\text{SCD}_a$, and SSL+$\text{SCD}_b$, achieve lower EERs than their matched-condition counterparts. In contrast, systems trained on A5 and evaluated on A19 consistently perform worse than the corresponding matched A19 systems. These results indicate that transferability between the two datasets is asymmetric and strongly influenced by differences in spoofing diversity and recording conditions. We believe that this asymmetric behavior stems from intrinsic differences between the ASVspoof 2019 LA and ASVspoof~5 datasets. The latter contains substantially more diverse and challenging spoofing conditions, including greater variability in recording environments, acoustic channels, and attacks. As a result, systems trained on ASVspoof~5 may become more specialized to its complex conditions, reducing their generalization capability to the comparatively more homogeneous ASVspoof 2019 LA dataset.

Another important observation is that $\text{SCD}_a$ and $\text{SCD}_b$ exhibit substantially better cross-dataset robustness than the original SCD representation, particularly in the A5$\rightarrow$A19 setting, where SSL+SCD experiences severe performance degradation. This suggests that the refinements introduced in $\text{SCD}_a$ and $\text{SCD}_b$ improve the robustness and transferability of cyclostationary spoofing cues under dataset mismatch conditions. Furthermore, although conventional handcrafted features such as LFCC and Mel-spectrogram also achieve strong cross-dataset performance, the consistently competitive results of SSL+$\text{SCD}_a$ and SSL+$\text{SCD}_b$ indicate that the proposed SC-based representations provide useful complementary information to SSL embeddings not only under matched conditions, but also in challenging cross-dataset scenarios.

\section{Discussion}
The results highlight several key observations.

\begin{itemize}[label={}, leftmargin=0pt, itemsep=0pt, topsep=0pt]
    \item \textbf{SSL dominance and generalization:} The SSL-only CM consistently yields the best performance across all datasets when compared with other single systems, namely SE-Res2Net50 and SSL-SE-Res2Net50. SSL-SE-Res2Net50 is inferior to the SSL-only system independent of the dataset. This supports the use of a lightweight back-end for SSL features.
    \item \textbf{Dataset-dependent behavior of acoustic front-ends:} For SE-Res2Net50, the relative ranking of baseline spectral and spectral correlation features varies across datasets, indicating sensitivity to channel/codec variability (ASVspoof 2021) and adversarial perturbations (ASVspoof 5).
    \item \textbf{Effectiveness of spectral correlation features:} The ordering of proposed spectral correlation features using SE-Res2Net50 back-end is dataset-dependent: on ASVspoof 2019, $\text{SCD}_b < \text{SCD} < \text{SCD}_a$, whereas on ASVspoof 2021 and ASVspoof 5 it becomes $\text{SCD}_b < \text{SCD}_a < \text{SCD}$ (for both EER and minDCF). Compared to baseline acoustic features, Mel-spectrogram is strongest on ASVspoof 2019, while $\text{SCD}_b$ is most competitive on ASVspoof 2021 and LFCC is superior on ASVspoof 5 for both EER and minDCF.
    \item \textbf{Fusion is advantageous across a wide range of conditions:} On ASVspoof 2019, embedding fusion yields large gains, and concatenation is consistently best; SSL+$\text{SCD}_b$ is the best performing configuration. On ASVspoof 2021, fusion typically does not improve on SSL-only, indicating limited complementarity under DF conditions, although it improves substantially over SE-Res2Net50 alone and $\text{SSL+}\text{SCD}_a$ is the best-performning combination among the fusion systems. On ASVspoof 5, fusion depends on the subset: SSL-only is best on the development set, while SSL+SCD is best on the evaluation set, with strong gains for adversarial attacks, suggesting conventional SCD captures complementary attack-relevant cues. Collectively, these findings suggest that higher-order cyclostationary information captured by SCD is complementary to, rather than redundant with, modern SSL representations, providing a signal-processing explanation for the effectiveness of the proposed hybrid systems across different datasets and architectures.
    \item \textbf{Impact of fusion strategy complexity:} The additional experiments on cross-attentive fusion and windowed binary gating further highlight the trade-off between fusion-model complexity and generalization capability. Although cross-attentive fusion achieved lower development-set EERs than the concatenation-based approach for several configurations, these gains did not consistently transfer to the evaluation set, where concatenation-based fusion remained superior overall.
    \item \textbf{Generalization across datasets:} The cross-dataset experiments provide additional evidence regarding the robustness of the proposed fusion framework under severe domain mismatch conditions. In particular, the proposed $\textnormal{SCD}_a$ and $\textnormal{SCD}_b$ representations consistently improved generalization compared with the SSL-only baseline in both studied configurations.
    \item \textbf{Performance--efficiency trade-off:} The inference-time analysis also demonstrates that the proposed concatenation-based SSL+$\textnormal{SCD}$ framework achieves a favorable performance and computational-cost balance. Although embedding fusion nearly doubles the inference latency compared with the SSL-only system, both systems remain substantially faster than real time.
\end{itemize}

Although the proposed SCD-based representations provide complementary information to SSL embeddings, their benefits are not uniform across all datasets, attack types, and evaluation conditions. Therefore, SCD should not be interpreted as a universal replacement for SSL-based representations or conventional acoustic front ends. From a deployment perspective, the use of SCD introduces additional feature-extraction and fusion overhead, and SSL-only systems may remain preferable in highly resource-constrained or latency-critical scenarios. The main value of SCD lies instead in providing an interpretable signal-processing representation that captures higher-order cyclostationary structure not explicitly modeled by conventional time-frequency features or current SSL pretraining objectives. Future work should therefore focus on more efficient SCD computation such as attention-based weighted averaging of frame-level correlations and earlier-stage fusion mechanisms that can better exploit frame-level interactions between cyclostationary and SSL representations.

In terms of generalization capability, spectral correlation-based features consistently exhibit smaller performance degradation when transitioning from the development set to the evaluation set across all datasets, indicating improved robustness.

\begin{figure}[!h]
\vspace{-3mm}
\centerline{\includegraphics[scale=0.14]{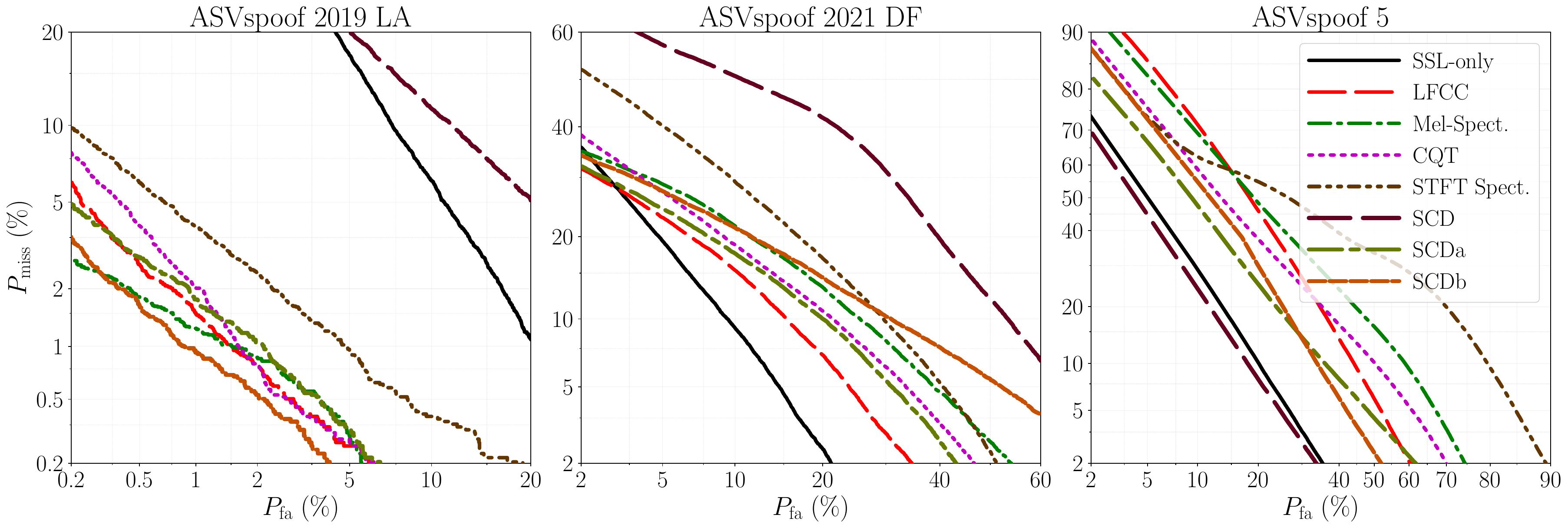}}
\vspace{-3mm}
\caption{DET Curves for ASVspoof 2019 LA, ASVspoof 2021 DF and ASVspoof 5 evaluation sets for the SSL-only (solid blue) and the concatenation-based fusion systems with different acoustic features.}
\vspace{-1mm}
\label{fig:det_plots}
\end{figure}

Although the impact of the proposed spectral correlation features are evident in terms of both metrics, these reflect system performance at a single operating point. To gain further insight into system behavior across a wider range of operating points, Fig.~\ref{fig:det_plots} presents the DET curves for the SSL-only system and the embedding fusion-based system using various acoustic features across the three datasets. The DET curves show that for ASVspoof 2019, the $\text{SSL+SCD}_b$ embedding fusion consistently outperforms other combinations. For ASVspoof 2021, the SSL-only system achieves the best performance, followed by SSL+LFCC and SSL+SCD$_a$. Finally, for ASVspoof 5, the SSL+SCD configuration systematically outperforms other combinations as well as the SSL-only system.

\begin{figure}[t]
\vspace{-3mm}
    \centerline{\includegraphics[scale=0.21]{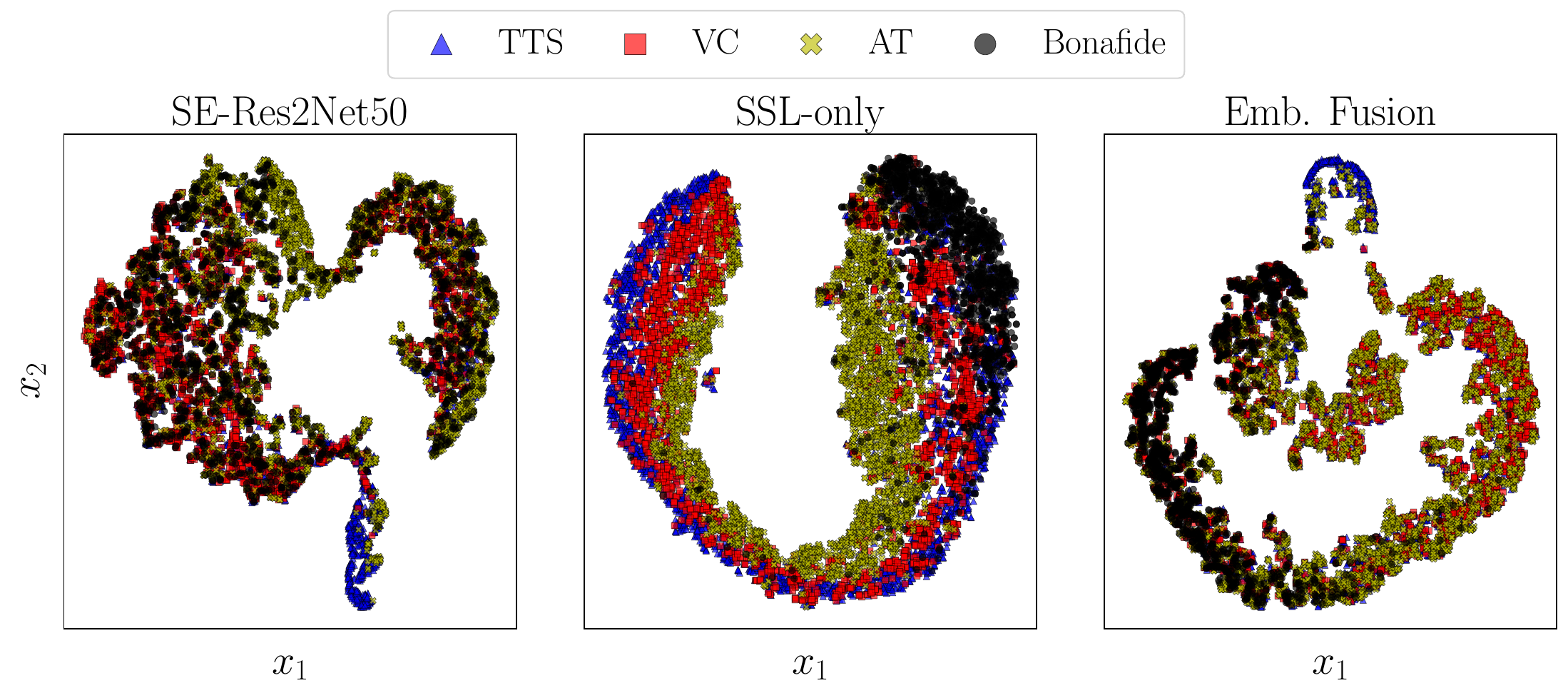}}
    \vspace{-3mm}
    \caption{t-SNE visualizations of embeddings on the ASVspoof 5 evaluation set from the SE-Res2Net50 (left), SSL-only (middle), and embedding fusion (right) systems. SE-Res2Net50 and the embedding fusion system utilize conventional SCD features, whereas the SSL-only system relies solely on self-supervised representations. Points are colored by class labels (bonafide, text-to-speech (TTS), voice conversion (VC), and adversarial attacks (AT)).}
    \label{fig:tsne_cm}
    \vspace{-3mm}
\end{figure}

Finally, Fig.~\ref{fig:tsne_cm} illustrates t-SNE projections of embeddings extracted from SE-Res2Net50, SSL-only and embedding fusion-based CM systems using the ASVspoof 5 evaluation set. SSL-only shows limited separation between bonafide and spoofed speech across attack types, whereas the systems leveraging conventional SCD features—namely SE-Res2Net50 and the embedding fusion approach—demonstrate more structured and compact embedding distributions. In particular, the fusion system produces clearer clustering, consistent with complementary information from combining SSL representations with conventional SCD embeddings.

\section{Conclusion}
This work investigated the use of cyclostationary signal processing for speech deepfake detection (SDD) through spectral correlation density (SCD) representations. Unlike conventional time–frequency features, SCD captures higher-order spectral correlations associated with the periodic structure of speech. Rather than replacing modern learned representations, the proposed SCD features were investigated as a complementary source of information to SSL embeddings.

Although recent advances in SDD have largely reduced the emphasis on handcrafted features in favor of SSL representations, the experimental results demonstrate that spectral correlation-based features remain highly relevant. Experiments on ASVspoof 2019 LA, ASVspoof 2021 DF, and the most recent and challenging ASVspoof 5 datasets, showed that SCD features achieve competitive standalone performance and consistently improve SSL-based systems through embedding-level fusion. The observed improvements indicate that higher-order spectral correlation features convey complementary spoofing cues that are not fully captured by conventional acoustic features or SSL-derived representations.

Future work will investigate more advanced fusion mechanisms between SSL and spectral-correlation-based representations, including attention-based and adaptive gating strategies. In particular, architectures performing fusion before utterance-level pooling may enable more effective modeling of frame-level interactions between SSL and cyclostationary representations, potentially improving the exploitation of complementary temporal and spectral-correlation cues while maintaining strong generalization capability. Future studies will also investigate multilingual and cross-domain evaluation settings, as well as the impact of different sampling frequencies and related SCD parameter optimization.

\bibliographystyle{IEEEtran}
\bibliography{References}

\iffalse
\newpage

\section{Biography Section}
If you have an EPS/PDF photo (graphicx package needed), extra braces are
 needed around the contents of the optional argument to biography to prevent
 the LaTeX parser from getting confused when it sees the complicated
 $\backslash${\tt{includegraphics}} command within an optional argument. (You can create
 your own custom macro containing the $\backslash${\tt{includegraphics}} command to make things
 simpler here.)
 
\vspace{11pt}

\bf{If you include a photo:}\vspace{-33pt}
\begin{IEEEbiography}[{\includegraphics[width=1in,height=1.25in,clip,keepaspectratio]{fig1}}]{Michael Shell}
Use $\backslash${\tt{begin\{IEEEbiography\}}} and then for the 1st argument use $\backslash${\tt{includegraphics}} to declare and link the author photo.
Use the author name as the 3rd argument followed by the biography text.
\end{IEEEbiography}

\vspace{11pt}

\bf{If you will not include a photo:}\vspace{-33pt}
\begin{IEEEbiographynophoto}{John Doe}
Use $\backslash${\tt{begin\{IEEEbiographynophoto\}}} and the author name as the argument followed by the biography text.
\end{IEEEbiographynophoto}

\vfill
\fi

\end{document}